\documentclass[journal]{IEEEtran}
\newtheorem{definition}{Definition}
\newtheorem{theorem}{Theorem}
\newenvironment{proof}{\begin{IEEEproof}}{\end{IEEEproof}}
\usepackage{amssymb}

\usepackage{multirow}  
\usepackage{hyperref}

\usepackage{amsmath}
\usepackage{algorithmic}
\usepackage{algorithm}
\usepackage{multirow}
\usepackage{array}
\usepackage{url}

\ifCLASSINFOpdf
\usepackage[pdftex]{graphicx}
\DeclareGraphicsExtensions{.pdf,.jpeg,.png}
\else
\usepackage[dvips]{graphicx}
\DeclareGraphicsExtensions{.eps}
\fi

\ifCLASSINFOpdf
\else
\fi

\begin{document}

\title{Explainable Fuzzy Utility Mining on Sequences}

\author{Wensheng Gan,~\IEEEmembership{Member,~IEEE,}
	Zilin Du,~\IEEEmembership{Student Member,~IEEE,}  
	Weiping Ding,~\IEEEmembership{Senior Member,~IEEE,}
	\\  Chunkai Zhang,~\IEEEmembership{Member,~IEEE,}	
	and Han-Chieh Chao,~\IEEEmembership{Senior Member,~IEEE}

\thanks{This research was partially supported by the National Natural Science Foundation of China (Grant No. 62002136, 61976120), Natural Science Foundation of Guangdong Province, China (Grant No. 2020A1515010970), Natural Science Foundation of Jiangsu Province (Grant No. BK20191445), sponsored by Qing Lan Project of Jiangsu Province, and Shenzhen Research Council (Grant No. GJHZ20180928155209705). Manuscript received January 1, 2021; revised February 28, 2021. (Corresponding author: Chunkai Zhang, E-mail: ckzhang@hit.edu.cn)}

	\thanks{Wensheng Gan is with the College of Cyber Security, Jinan University, Guangzhou 510632, Guangdong, China; with the Department of Computer Science and Technology, Harbin Institute of Technology (Shenzhen), Shenzhen 518055, China; and with Guangdong Artificial Intelligence and Digital Economy Laboratory (Guangzhou), China.} 

	\thanks{Zilin Du and Chunkai Zhang are with Department of Computer Science and Technology, Harbin Institute of Technology (Shenzhen), Shenzhen 518055, China.} 

	\thanks{Weiping Ding is with the School of Information Science and Technology, Nantong University, Nantong 226019, Jiangsu, China.} 
	
	\thanks{Han-Chieh Chao is with the Department of Electrical Engineering, National Dong Hwa University, Hualien, Taiwan, R.O.C.} 

}

\markboth{IEEE Transactions on Fuzzy Systems, 2021}%
{Gan \MakeLowercase{\textit{et al.}}: Bare Demo of IEEEtran.cls for IEEE Journals}

\maketitle

\begin{abstract}

Fuzzy systems have good modeling capabilities in several data science scenarios, and can provide human-explainable intelligence models with explainability and interpretability. In contrast to transaction data, which have been extensively studied, sequence data are more common in real-life applications. To obtain a human-explainable data intelligence model for decision making, in this study, we investigate explainable fuzzy-theoretic utility mining on multi-sequences. Meanwhile, a more normative formulation of the problem of fuzzy utility mining on sequences is formulated. By exploring fuzzy set theory for utility mining, we propose a novel method termed pattern growth fuzzy utility mining (PGFUM) for mining fuzzy high-utility sequences with linguistic meaning. In the case of sequence data, PGFUM reflects the fuzzy quantity and utility regions of sequences. To improve the efficiency and feasibility of PGFUM, we develop two compressed data structures with explainable fuzziness. Furthermore, one existing and two new upper bounds on the explainable fuzzy utility of candidates are adopted in three proposed pruning strategies to substantially reduce the search space and thus expedite the mining process. Finally, the proposed PGFUM algorithm is compared with PFUS, which is the only currently available method for the same task, through extensive experimental evaluation. It is demonstrated that PGFUM achieves not only human-explainable mining results that contain the original nature of revealable intelligibility, but also high efficiency in terms of runtime and memory cost.

\end{abstract}


\IEEEpeerreviewmaketitle


\section{Introduction}  \label{sec:introduction}

\IEEEPARstart{W}{ith} the explosion of rich data, which are commonly seen on the Internet, e-commerce, World Wide Web, and social networks, their understanding, modeling, and mining is a critical and challenging task. Data mining has provided several useful models and algorithms for discovering interesting knowledge from databases. For example, association rule mining, including frequent pattern mining (FPM) \cite{agrawal1993mining}, fuzzy rule mining  \cite{hong2003fuzzy}, and sequential pattern mining (SPM) \cite{han2001prefixspan}, is important in data analytics. Although several efficient methods have been proposed to handle a large amount of rich data, they are ineffective and inefficient for discovering highly profitable patterns. This is because they perform the mining task without considering certain interesting measures such as utility. These challenges motivated a new mining framework termed utility-driven pattern mining (abbreviated as utility mining) \cite{gan2020survey}. In the framework of utility mining on itemset-based data (also termed high-utility itemset mining, HUIM) \cite{yao2004foundational} and utility mining on sequence-based data (also termed high-utility SPM, HUSPM) \cite{gan2020proum,yin2012uspan}, a subjective  measure (with respect to the concept of utility) and other objective measures are adopted to discover desired patterns. In these frameworks, various effective models and efficient algorithms \cite{gan2020proum,yin2012uspan} have been extensively studied. In addition, several novel pruning strategies and data structures have been designed to efficiently discover high-utility patterns under various constraints \cite{gan2020survey}.

Utility mining on sequence data is an emerging issue in knowledge discovery with wide real-life applications. It considers not only the utility but also the occurrence order of items in sequences. Therefore, by mining useful high-utility patterns, utility mining on sequence data can assist managers to discover potential interesting marketing knowledge and make appropriate decisions. In HUSPM, there are several high-performance algorithms. However, the application of utility-mining techniques involves certain challenges. Achieving human-explainable data mining and data intelligence is quite challenging \cite{arrieta2020explainable}. Most studies do not consider linguistic terms in fuzzy theory. More importantly, providing an explanation to the processed data and mining results is increasingly important to systems based on artificial intelligence (AI), particularly data mining models and algorithms.

Owing to the lack of transparency and linguistic meaning, most data intelligence systems cannot explain the data and discovered results in a human-friendly form. In general, a fuzzy set can express and discover meaningful information regarding the logic involved. For example, either on-demand explanations or a model description are more interpretable. To resolve these issues and challenges, explainable AI systems \cite{dovsilovic2018explainable} regarding internal data representations and decisions have been further studied. In recent years, explainable AI (XAI) \cite{arrieta2020explainable,samek2017explainable,gunning2017explainable} has attracted wide attention. In general, XAI usually includes explainable data-mining techniques \cite{hong2003fuzzy,hong2016survey,bao2018knowledge}, explainable neural networks \cite{arrieta2020explainable}, and fuzzy systems \cite{cao2017inherent,zhao2019fuzzy,tirkolaee2020fuzzy}. The potential explanation should be understandable not only by experts in the target domain, but also by end-users. Currently, XAI is a new trend that provides explanations of the intelligent decisions of AI algorithms, and fuzzy theory is useful in XAI. 

To summarize, traditional utility-driven sequential pattern mining algorithms lack explainability for complex systems. With the growing use of utility mining and fuzzy systems, explainable HUSPM addresses the need for explainability. To the best of our knowledge, only projection-based fuzzy utility sequential (PFUS) pattern mining  \cite{lan2013mining} is related to the task of fuzzy utility mining (FUM) on sequences. However, this method cannot efficiently handle real-life databases because it requires intensive mathematical computations. In this study, we address the problem of FUM on sequence data for decision making with suitable explainability. By applying the concept of fuzzy sets \cite{hong2003fuzzy,zadeh1988fuzzy} to utility mining, we can not only obtain a human-explainable data-driven intelligence model, but also significantly improve its mining efficiency. The major contributions of this study can be summarized as follows:

\begin{itemize}
	\item 	We provide a more normative formulation of the problem of FUM on sequences. We propose a novel explainable method: pattern-growth fuzzy utility mining (PGFUM) on multi-sequences with linguistic meaning. 
	
	\item  To store rich information from sequence data, we develop two compressed data structures with explainable fuzziness: fuzzy matrix set and fuzzy utility chain. 
	
	\item  One existing and two new upper bounds on the explainable fuzzy utility of candidates are adopted in three proposed pruning strategies to substantially reduce the search space and thus expedite the mining process.
	
	\item  Experiments on popular benchmarks, both synthetic and real datasets, demonstrate that PGFUM can efficiently discover high-fuzzy-utility sequential patterns (HFUSPs). Human-explainable mining results contain the original nature of the revealable intelligibility, and PGFUM can achieve better performance than the state-of-the-art algorithm.
\end{itemize}

The remainder of this paper is organized as follows. Related work is briefly reviewed in Section \ref{sec:relatedwork}. Then, the problem of FUM on sequences is formulated in Section \ref{sec:preliminaries} alongside related definitions. In Section \ref{sec:method}, we present the proposed PGFUM algorithm with several detailed strategies and data structures. The experimental results are evaluated in Section \ref{sec:experiments}. Finally, Section \ref{sec:conclusion} concludes this paper and discusses several future work.

\section{Review of Related Work}  \label{sec:relatedwork}

\subsection{Utility Mining on Sequence Data}

Frequent pattern mining has been extensively investigated since it was first proposed as large itemset mining \cite{agrawal1994fast}. It is now usually termed frequent itemset mining (FIM), and a large number of methods, such as FP-Growth \cite{han2004mining}, have been developed to improve the performance of discovering frequent itemsets as desired patterns. Generally, FIM is regarded as a common knowledge discovery approach with a wide range of applications, such as image classification and e-learning, as transaction data are common in real life. It has been extended to SPM \cite{zaki2001spade,van2018mining,fournier2017survey}, which focuses on extracting frequent subsequences (i.e., sequential patterns) from sequence data. Some researchers have applied SPM to genome and clickstream analysis \cite{van2018mining,kieu2017mining}. The aforementioned FPM framework discovers frequency patterns that may indicate user interests. Utility mining is a new computing framework that is different from frequency-based FPM \cite{agrawal1993mining} and SPM models \cite{kieu2017mining,fournier2017survey}. In general, utility-driven mining incorporates utility theory from Economics into the mining process. It focuses on the benefit of the mining results that contribute to the overall benefit, maximizing utility. Inspired by FPM, most modern utility mining algorithms are Apriori-like \cite{liu2005two} or tree-based methods \cite{tseng2010up}. To improve mining efficiency, vertical data-structure algorithms have been developed, such as HUI-Miner \cite{liu2012mining}, HUOPM \cite{gan2019huopm}, EFIM \cite{zida2015efim}, and others \cite{lin2017fdhup,nguyen2019mining}. Most of these algorithms are related to transactional but not sequence data; the latter are more complicated but commonly seen in real life. Motivated by sequence data that usually contain rich information (e.g., timestamp, quantity, or dimension), HUSPM \cite{gan2020proum,yin2012uspan,wang2016efficiently,gan2021utility} has been studied. To date,  there have been a few utility mining algorithms for handling different types of sequence data. For example, USpan \cite{yin2012uspan}, ProUM \cite{gan2020proum}, and HUSP-ULL \cite{gan2020fast} have been proposed to handle normal sequence data, and USPT \cite{gan2020utility} is a novel algorithm for utility mining across multi-sequences with individualized thresholds. Recent advances in utility mining on different types of data have been reviewed in detail in Ref. \cite{gan2020survey}. 

However, traditional utility-driven sequential pattern mining algorithms lack explainability for complex systems. With the growing use of utility mining, sufficient explainability of the processed data as well as mining results is required. The concept of fuzzy sets \cite{hong2003fuzzy,zadeh1988fuzzy,khalilpourazari2019robust2} has the benefits of simplicity and comprehensibility to human reasoning \cite{arrieta2020explainable}. Lan \textit{et al.} \cite{lan2013mining} first proposed the PFUS algorithm utilizing a projection technique for high efficiency. To the best of our knowledge, this is the only fuzzy-theoretic approach to HUSPM. With the consideration of linguistic terms in fuzzy utility mining, PFUS has more explainability and interpretability than traditional HUSPM models. Unfortunately, as an AprioriAll-like method \cite{agrawal1995mining}, PFUS suffers from serious performance bottlenecks, although it uses an upper bound on sequence utility to prune the search space. Achieving satisfactory execution efficiency for fuzzy-theoretic utility mining on sequence data remains a challenging issue.

\subsection{Fuzzy-logic-based Pattern Mining}

Early studies on association rule mining (ARM) \cite{agrawal1993mining} primarily considered the co-occurrence of items in transactions. However, the relationships (e.g., occurred quantities) between items were not considered. This is problematic, as for instance, frequent patterns based on Boolean values are not meaningful because quantitative information is ignored. To address this issue, Srikant \textit{et al.} \cite{srikant1996mining} first introduced quantitative ARM and proposed an Apriori-like mining algorithm. However, as mentioned previously, all these models and algorithms lack simplicity and comprehensibility. Therefore, fuzzy pattern mining \cite{hong2003fuzzy,kuok1998mining} was proposed by incorporating fuzzy set theory into traditional data mining. A number of studies have demonstrated that fuzzy sets and systems \cite{hong2003fuzzy,zadeh1988fuzzy,khalilpourazari2019robust1} possess excellent fuzzy modeling capabilities in several data science scenarios, as they consist of human reasoning and decision models. Fuzzy set theory has been widely used in several real-life applications because of its simplicity and explainability to human reasoning \cite{hong2016survey}.

To evaluate the count of fuzzy association rules in a transaction database, Hong \textit{et al.} \cite{hong2003fuzzy} proposed the fuzzy minimum operator in fuzzy set theory. This is different from the calculation function in the study by Kuok \textit{et al.} \cite{kuok1998mining}. Subsequently, several more efficient fuzzy data mining approaches were proposed to improve mining efficiency, and details can be found in  Ref. \cite{hong2016survey}. Wang \textit{et al.} \cite{wang2009fuzzy} first integrated fuzzy theory with utility mining, that is, FUM, to handle quantitative databases, and proposed fuzzy high-utility itemset mining, also termed FUM, on transactions. The fuzzy utility function in FUM considers both the internal and the weighted external utility to evaluate the fuzzy utility of an itemset. In FUM on transactions \cite{wang2009fuzzy,lan2015fuzzy}, the mining results are fuzzy high-utility itemsets that satisfy a user-specified minimum fuzzy utility threshold. To date, the problem of fuzzy-driven data intelligence has been extensively studied and applied to various real-world applications \cite{hong2016survey}, such as fuzzy association mining \cite{hong2003fuzzy}, fuzzy sequential pattern analysis \cite{lan2013mining}, fuzzy classification \cite{zhao2019fuzzy}, and other fuzzy systems \cite{bao2018knowledge,cao2017inherent,goli2021fuzzy}. As mentioned previously, this motivates our exploration of two issues:  utility mining on quantitative sequence data and fuzzy-driven pattern mining. In addition, developing effective and efficient FUM approaches for handling sequence data to support interpreting and explaining mining results is challenging.

\section{Preliminaries and Problem Formulation}  \label{sec:preliminaries}

Herein, we briefly present the basic concepts and notations of utility and fuzzy utility, as summarized in TABLE \ref{notations}, based on related definitions from prior studies \cite{lan2013mining,gan2020fast,lan2015fuzzy}. In addition, we provide a more normative formulation of the problem of FUM on sequences.

\begin{table}[!ht]
	\caption{Summary of symbols and their meanings}
	\label{notations}
	\centering
	\begin{tabular}{|c||p{6cm}|}
		\hline
		\textbf{Symbol}	    & \makebox[6cm][c]{\textbf{Definition}} \\ \hline 
		$I$	& A finite set of items may appear. \\ \hline
		$\textit{FS}$	& A $f$-sequence. \\ \hline
		$\textit{QS}$	& A $q$-sequence. \\ \hline
		$D$	& A $q$-sequence database. \\ \hline
		$u(D)$	& The overall utility of the $q$-sequence database $D$. \\ \hline
		$q(i,\textit{QX})$	& The internal utility of item $i$ in the $q$-itemset \textit{QX}. \\ \hline
		$p(i)$	& The user-defined external utility of item $i$. \\ \hline
		$\textit{\textit{fu}(\textit{FS})}$	& The fuzzy utility of the $f$-sequence \textit{FS}. \\ \hline
		FUM	& Fuzzy utility mining. \\ \hline
		HFUSP & High-fuzzy-utility sequential pattern. \\ \hline
		FE-tree	& A fuzzy extension tree. \\ \hline
		\textit{MFUI} & The maximum fuzzy utility of an item. \\ \hline 
		\textit{MRFU} & The maximum rest fuzzy utility value. \\ \hline 
		\textit{HFSUUB}	& The high fuzzy sequence-utility upper bound. \\ \hline
		\textit{SDFU}	& The upper bound sequence descendant fuzzy utility. \\ \hline
		\textit{EIFU}	& The upper bound extension item fuzzy utility. \\ \hline
	\end{tabular}
\end{table}

\subsection{Preliminaries Regarding Utility}

Let $I$ = $\{i_{1}$, $i_{2}$, $\cdots$, $i_{N}\}$ be a finite set of $N$ distinct elements, each of which is called an item. Given a nonempty set $X$ that satisfies $X \subseteq I$, we say that $X$ is an itemset. The length of $X$ is defined as the number of items in $X$ and is denoted as $|X|$. Given sequential relations in a series of itemsets, $S$ = $<$$X_{1}$, $X_{2}$, $\cdots$, $X_{n}$$>$ is an ordered list of itemsets called sequence, where $X_{k} \subseteq I$ for $1 \leq k \leq n$. We define the length of $S$ as $|S|$ = $\sum_{k = 1}^{n}|X_{k}|$, whereas the size of $S$ is $n$. Given a sequence $T$ = $<$$Y_{1}$, $Y_{2}$, $\cdots$, $Y_{m}$$>$, $T$ is called a subsequence of $S$ if and only if $m \leq n$ and there exist $m$ positive integers $1 \leq $ $ k_{1}$ $< k_{2} $ $< \cdots < k_{m}$ $ \leq n$ so that $Y_{v} \subseteq X_{k_{v}}$ for all $1 \leq v \leq m$. For example, $t$ = \{\textit{c} \textit{d}\} is an itemset with length $|t|$ equal to 2. Given two sequences $s_{1}$ = $<$\{\textit{a} \textit{b}\}, \{\textit{e}\}$>$, and $s_{2}$ = $<$\{\textit{a} \textit{b} \textit{f}\}, \{\textit{b} \textit{e} \textit{f}\}, \{\textit{c}\}$>$, $s_{1}$ is a subsequence of $s_{2}$. We have $|s_{1}|$ = 3 because $s_{1}$ consists of three items.

\begin{definition}
	A quantitative itemset $\textit{QX}$ = \{(\textit{$i_{1}$}:$q_{1}$) (\textit{$i_{2}$}:$q_{2}$) $\cdots$ (\textit{$i_{m}$}:$q_{m}$)\} is a finite set with $m$ elements, each of which is an ordered tuple called quantitative item. In a quantitative item (\textit{i}:$q$), the first symbol represents an item \textit{i} satisfying $\textit{i} \in I$, and $q$ is a positive integer called internal utility (e.g., representing a quantity bought once) of \textit{i}. $\textit{QS}$ = $<$$\textit{QX}_{1}$, $\textit{QX}_{2}$, $\cdots$, $\textit{QX}_{n}$$>$ is called a quantitative sequence, where  $\textit{QX}_{k}$ is a quantitative itemset for $1 \leq k \leq n$. The terms quantitative item/quantitative itemset/quantitative sequence are commonly abbreviated as $q$-item/$q$-itemset/$q$-sequence. 
\end{definition}

We define the length of a $q$-itemset $\textit{QX}$, denoted as $|\textit{QX}|$, to be the number of $q$-items in $\textit{QX}$. Similarly, the length of a $q$-sequence $\textit{QS}$ = $<$$\textit{QX}_{1}$, $\textit{QX}_{2}$, $\cdots$, $\textit{QX}_{n}$$>$ is denoted as $|\textit{QS}|$, and is defined as $|\textit{QS}|$ = $\sum_{k = 1}^{n}|\textit{QX}_{k}|$, where $1 \leq k \leq n$.

\begin{table}[!t]
	\centering
	\caption{An running example of $q$-sequence database}
	\label{table1}
	\begin{tabular}{|c||c|}  
		\hline 
		\textbf{SID} & \textbf{$q$-sequence} \\
		\hline  
		\(\textit{QS}_{1}\) & $<$\{(\textit{b}:2) (\textit{d}:3)\}, \{(\textit{a}:3) (\textit{e}:2)\}, \{(\textit{b}:1) (\textit{c}:4) (\textit{e}:3)\}$>$ \\ 
		\hline
		\(\textit{QS}_{2}\) & $<$\{(\textit{a}:3) (\textit{c}:4)\}, \{(\textit{a}:4) (\textit{d}:1)\}, \{(\textit{a}:4) (\textit{c}:2) (\textit{e}:1)\}, \{(\textit{d}:5)\}$>$ \\  
		\hline  
		\(\textit{QS}_{3}\) & $<$\{(\textit{a}:1) (\textit{c}:2)\}, \{(\textit{e}:2)\}, \{(\textit{a}:2) (\textit{d}:3) (\textit{e}:1)\}$>$ \\
		\hline  
		\(\textit{QS}_{4}\) & $<$\{(\textit{a}:1)\}, \{(\textit{a}:3) (\textit{c}:1)\} \{(\textit{d}:4)\} \{(\textit{f}:1)\}$>$ \\
		\hline
	\end{tabular}
\end{table}

\begin{table}
	\caption{A utility table}
	\label{table2}
	\centering
	\begin{tabular}{|c||c||c||c||c||c||c|}
		\hline
		\textbf{Item}	    & \textit{a}	& \textit{b}	& \textit{c}	& \textit{d}	& \textit{e}	& \textit{f} \\ \hline 
		\textbf{External utility}	& \$2 & \$1	& \$3 & \$4 & \$2 & \$5 \\ \hline
	\end{tabular}
\end{table}

\begin{definition}
	In FUM on sequences, the mining object is a $q$-sequence database $D$, which is a set of $q$-sequences $D$ = \{$\textit{QS}_{1}$, $\textit{QS}_{2}$, $\cdots$, $\textit{QS}_{n}$\}, where each $q$-sequence, also called sequential transaction, has a unique identifier \textit{SID}. Moreover, each item type is associated with a positive integer called external utility (e.g., representing the unit profit).
\end{definition}

A running example is shown in TABLE \ref{table1}. In the given $q$-sequence database, there are four $q$-sequences and six item types, the external utility values of which can be found in TABLE \ref{table2}. The first $q$-sequence $\textit{QS}_{1}$ consists of three $q$-itemsets, and we have $|\textit{QS}_{1}|$ = 2 + 2 + 3 = 7.

Given a $q$-itemset $\textit{QX}$ = \{(\textit{$i_{1}$}:$q_{1}$) (\textit{$i_{2}$}:$q_{2}$) $\cdots$ (\textit{$i_{m}$}:$q_{m}$)\} with length $m$, the utility of an item $i_{k}$ in $\textit{QX}$ is defined as $u(i_{k},\textit{QX})$ = $q(i_{k},\textit{QX})$ $ \times $ $p(i_{k})$, where $q(i_{k},\textit{QX})$ is the internal utility of item $i_{k}$ in $X$, and $p(i_{k})$ is the external utility of $i_{k}$ for $1 \leq k \leq m$. Let $u(\textit{QX})$ represent the utility of $\textit{QX}$, which can be obtained by $u(\textit{QX})$ = $\sum_{k = 1}^{m}u(i_{k},\textit{QX})$ for $1 \leq k \leq m$. Then, the utility of a $q$-sequence $\textit{QS}$, denoted as $u(\textit{QS})$, may be obtained analogously, that is, as the sum of the utilities of $q$-itemsets in $\textit{QS}$. Moreover, the utility of $D$ is denoted by $u(D)$ and is defined as the overall utility of the $q$-sequences in $D$. 

In the given running example, the utility of item \textit{d} in the \textit{1st} itemset in $\textit{QS}_{1}$ is $u$(\textit{b}$,  $\{(\textit{b}:2) (\textit{d}:3)\}) = 3 $\times$ \$4 = \$12. Furthermore, $u(\textit{QS}_{1})$ = \$14 + \$10 + \$19 = \$43 and $u(D)$ = \$43 + \$66 + \$30 + \$32 = \$171, as shown in TABLE \ref{table1}.

\subsection{Preliminaries Regarding Explainable Fuzzy Utility}

Background material regarding the economic concept of utility can be found in Ref. \cite{gan2020proum,yin2012uspan,wang2016efficiently}. Herein, based on the aforementioned preliminaries, we incorporate fuzziness into the concept of utility, and introduce several important definitions regarding fuzzy utility with explainability. 

\begin{definition}
	According to fuzzy set theory \cite{zadeh1965fuzzy}, assuming that the $m$-th item in the $n$-th $q$-sequence in $D$ is $i_{\textit{mn}}$, the fuzzy set $\textit{fz}_\textit{mn}$ of the explainable utility value of $i_{\textit{mn}}$ (i.e., $u(i_{\textit{mn}})$) can be represented by a given explainable membership function for the item $i_{\textit{mn}}$ as
	\begin{equation}
	\textit{fz}_\textit{mn}\ =\ (\frac{\textit{fz}_\textit{mn1}}{R_{m1}}+\frac{\textit{fz}_\textit{mn2}}{R_{m2}}+\cdots+\frac{\textit{fz}_\textit{mnh}}{R_{mh}})\ (\textit{fz}_\textit{mnk} \in [0,1])
	\end{equation}
	where $h$ is the number of regions of $i_{\textit{mn}}$, $\textit{fz}_\textit{mnk}$ is the explainable membership value (also known as fuzzy value and membership degree) of $i_{\textit{mn}}$ in the $k$-th region, $R_{mh}$ is the $k$-th region of $i_{m}$ for $1 \leq k \leq h$, and $D$ is a $q$-sequence database.
\end{definition}

\begin{figure}[htbp]
	\centering
	\includegraphics[scale=0.25]{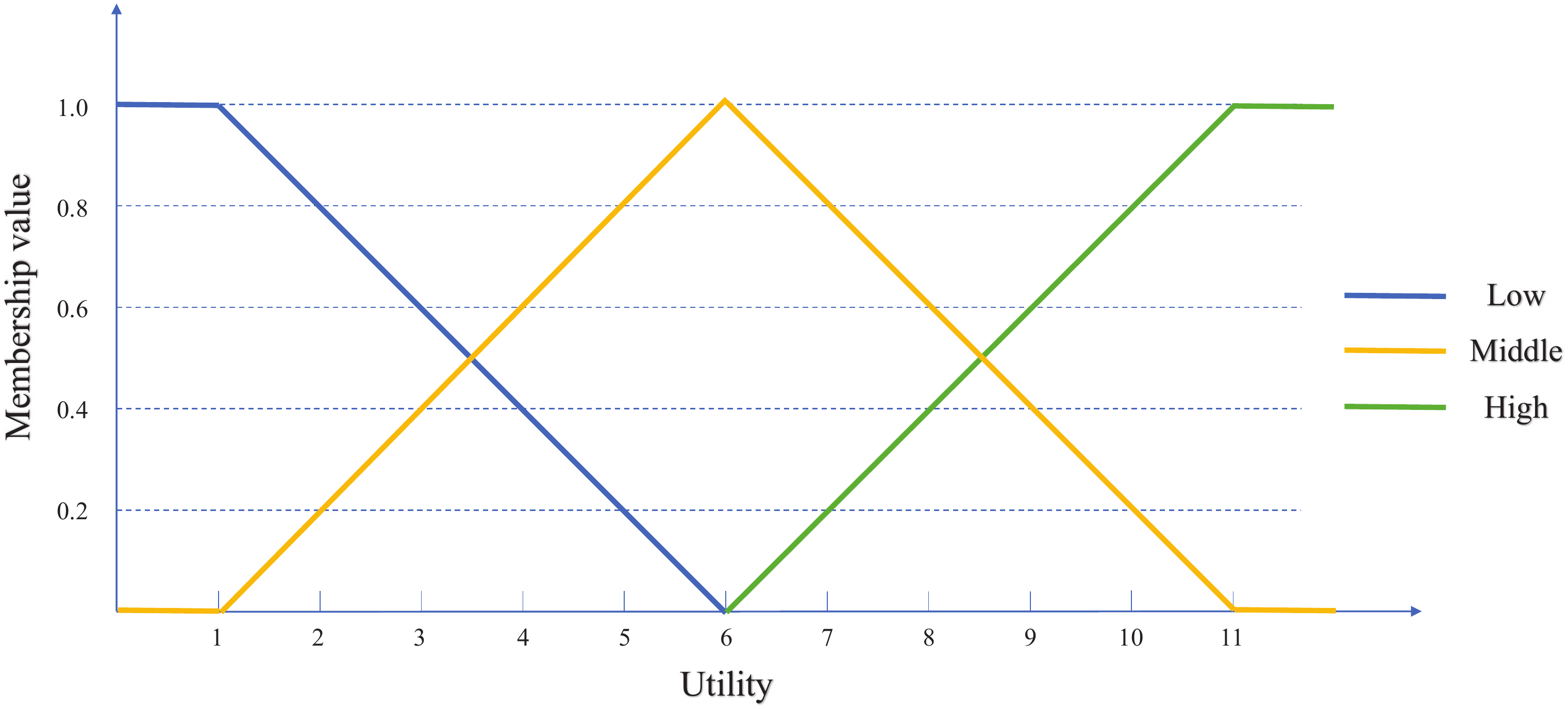}
	\caption{Membership function for the running example.}
	\label{membership_function}
\end{figure}

We note that the fuzzy set is the set of regions with their membership degrees, which are obtained from utilities by a membership function in the problem of FUM on sequences. To facilitate the following discussion, we introduce a membership function suitable for the running example above, as shown in Fig. \ref{membership_function}. The utilities in the running example distribute over [1,20]; thus, we consider an appropriate membership function curve to ensure that few membership values in different regions are equal to 0. The membership function used in the proposed method is user-defined according to a priori knowledge. Here, we take the item \textit{b} in the \textit{1st} $q$-itemset in $\textit{QS}_{1}$ as an example. The utility can be calculated as \$2 (2 $\times$ \$1), and it can be transformed into the fuzzy set  $(\frac{0.8}{\textit{b.Low}}+\frac{0.2}{\textit{b.Middle}})$. The region \textit{b.High} is omitted because the corresponding membership value is equal to 0. Moreover, $QS_{1}$ can be converted to $<$\{($\frac{0.8}{\textit{b.Low}}+\frac{0.2}{\textit{b.Middle}}$) ($\frac{1.0}{\textit{d.High}}$)\}, \{($\frac{1.0}{\textit{a.Middle}}$) ($\frac{0.4}{\textit{e.Low}}+\frac{0.6}{\textit{e.Middle}}$)\} \{($\frac{1.0}{\textit{b.Low}}$) ($\frac{1.0}{\textit{c.High}}$) ($\frac{1.0}{\textit{e.Middle}}$)\}$>$ by the same membership function. The other $q$-sequences can be processed similarly. 

\begin{definition}
	A fuzzy item is a tuple of the form (\textit{i}:$\textit{Region}_{k}$), where $\textit{i} \in I$, and $\textit{Region}_{k}$ is the $k$-th region of \textit{i}. 
	A fuzzy itemset $\textit{FX}$ = \{(\textit{$i_{1}$}:$\textit{Region}_{k_{1}}$) (\textit{$i_{2}$}:$\textit{Region}_{k_{2}}$) $\cdots$ (\textit{$i_{m}$}:$\textit{Region}_{k_{m}}$)\} is a finite set with $m$ fuzzy items. In addition, $\textit{FS}$ = $<$$\textit{FX}_{1}$, $\textit{FX}_{2}$, $\cdots$, $\textit{FX}_{n}$$>$ is an ordered list of a series of fuzzy itemsets, called a fuzzy sequence, where $\textit{FX}_{j}$ is a fuzzy itemset for $1 \leq j \leq n$.
\end{definition}

For convenience, "$f$-" is used to refer to an object associated with fuzziness in the remainder of this paper. In particular, the length of a $f$-sequence \textit{FS}, denoted as $|\textit{FS}|$, is the number of $f$-items in the $f$-sequence. An $f$-sequence with length $l$ can be represented as an $l$-sequence. 

For example, (\textit{a}:\textit{Low}) is an $f$-item, \{(\textit{a}:\textit{Low}) (\textit{b}:\textit{High})\} is an $f$-itemset, and $<$\{(\textit{e}:\textit{High})\}, \{(\textit{c}:\textit{Middle}) (\textit{b}:\textit{High})\}$>$ is an $f$-sequence with length 3 (i.e., 3-$f$-sequence). Without loss of generality, we assume that all items (e.g., $q$-items or $f$-items) in an itemset (e.g., $q$-itemset/$f$-itemset) are arranged in alphabetical order. 

\begin{definition}
	\label{fitem_utility}
	Given a $q$-sequence $\textit{QS}$ = $<$$\textit{QX}_{1}$, $\textit{QX}_{2}$, $\cdots$, $\textit{QX}_{n}$$>$, the fuzzy utility of an $f$-item (\textit{i}:$\textit{Region}_{k}$) in the $j$-th $q$-itemset in $\textit{QS}$ is defined as $\textit{fu}(i,\ j,\ \textit{QS})$ =  $q(i,X_{j})$ $\times$ $p(i)$ $\times$ $\textit{fz}_\textit{ik}$ = $u(i,X_{j})$ $\times$ $\textit{fz}_\textit{ik}$.
\end{definition}

For instance, in TABLE \ref{table1}, let us consider the $f$-item \textit{e} in the \textit{2nd} $q$-itemset in $\textit{QS}_{1}$. We have $u$((\textit{e}:\textit{Low}, 2, $\textit{QS}_{1}$)) = 2 $\times$ \$2 $\times$ 0.4 = \$1.6 and $u$((\textit{e}:\textit{Middle}, 2, $\textit{QS}_{1}$)) = 2 $\times$ \$2 $\times$ 0.6 = \$2.4. However, the fuzzy utility of (\textit{e}:\textit{High}) is equal to 0 because the membership value of \textit{e} in the \textit{3rd} region is 0.

\begin{definition}
	\label{contain}
	Given a $q$-itemset \textit{QX} = \{(\textit{$i_{1}$}:$q_{1}$) (\textit{$i_{2}$}:$q_{2}$) $\cdots$ (\textit{$i_{m}$}:$q_{n}$)\} and an $f$-sequence \textit{FX} = \{(\textit{$j_{1}$}:$\textit{Region}_{k_{1}}$) (\textit{$j_{2}$}:$\textit{Region}_{k_{2}}$) $\cdots$ (\textit{$j_{m}$}:$\textit{Region}_{k_{m}}$)\}, we say that \textit{QX} contains \textit{FX}, and use the notation $\textit{FX} \sqsubseteq \textit{QX}$, if there exist $m$ integers $ 1 \leq b_{1}$ $< b_{2} $ $< \cdots $ $< b_{m} \leq n$ such that \textit{$i_{b_{p}}$} = \textit{$j_{p}$} and $fz_{i_{b_{p}}k_{b_{p}}} > 0$ for $ 1 \leq p \leq m$, where $fz_{i_{b_{p}}k_{b_{p}}}$ is the membership value of $i_{b_{p}}$ in the $k_{b_{p}}$-th region. Let \textit{QS} = $<$$\textit{QX}_{1}$, $\textit{QX}_{2}$, $\cdots$, $\textit{QX}_{n}$$>$ be a $q$-sequence, and \textit{FS} = $<$$\textit{FX}_{1},\textit{FX}_{2},\cdots,\textit{FX}_{m}$$>$ be an $f$-sequence. If there exists an integer sequence $ 1 \leq$ $b_{1}$ $\leq b_{2}$ $\cdots \leq b_{m} \leq n$ such that $\textit{FX}_k \sqsubseteq \textit{QS}_{b_k}$ for $1 \leq k \leq m$, then \textit{QS} is said to contain \textit{FS}; this is denoted as $\textit{FS} \sqsubseteq \textit{QS}$.
\end{definition}

We have discussed that $QS_{1}$ can be converted to $<$\{($\frac{0.8}{\textit{b.Low}}+\frac{0.2}{\textit{b.Middle}}$) ($\frac{1.0}{\textit{d.High}}$)\}, \{($\frac{1.0}{\textit{a.Middle}}$) ($\frac{0.4}{\textit{e.Low}}+\frac{0.6}{\textit{e.Middle}}$)\} \{($\frac{1.0}{\textit{a.Low}}$) ($\frac{1.0}{\textit{c.High}}$) ($\frac{1.0}{\textit{e.Middle}}$)\}$>$ by the membership function shown in Fig. \ref{membership_function}. In the running example in TABLE \ref{table1}, the $f$-itemsets $\textit{FX}_{1}$ = \{(\textit{b}:\textit{Low}) (\textit{d}:\textit{High})\} and $\textit{FX}_{2}$ = \{(\textit{b}:\textit{Low})\} are both contained in the \textit{1st} $q$-itemset of $\textit{QS}_{1}$, whereas $\textit{FX}_{3}$ = \{(\textit{b}:\textit{High}) (\textit{d}:\textit{High})\} is not. Moreover, the $f$-sequence \textit{FS} = $<$\{(\textit{d}:\textit{High})\}, \{(\textit{a}:\textit{Middle}) (\textit{e}:\textit{Middle})\}$>$ is contained in \textit{QS}.

\begin{definition}
	Let \textit{QS} = $<$$\textit{QX}_{1}$, $\textit{QX}_{2}$, $\cdots$, $\textit{QX}_{n}$$>$ be a $q$-sequence, and \textit{FS} = $<$$\textit{FX}_{1}$, $\textit{FX}_{2}$, $\cdots$, $\textit{FX}_{m}$$>$ be an $f$-sequence. Assuming that, in Definition \ref{contain}, $\textit{FX} \sqsubseteq \textit{QX}$ and the integer sequence is $ 1 \leq k_{1} $ $\leq k_{2}$ $\cdots $ $\leq k_{m} \leq n$, we say that \textit{FS} has an instance in \textit{QS} at position $p$: $<$$k_{1}$, $k_{2}$, $\cdots$, $k_{m}$$>$.
\end{definition}

For instance, in TABLE \ref{table1}, an $f$-sequence $\textit{FS}_{1}$ = $<$\{(\textit{d}:\textit{High})\}, \{(\textit{c}:\textit{High})\}$>$ has an instance in $\textit{QS}_{1}$ at position $<$1,3$>$. We note that, in some cases, there may be more than one instances. For example, the $f$-sequence $\textit{FS}_{2}$ = $<$\{(\textit{b}:\textit{Low})\},  \{ (\textit{e}:\textit{Middle})\}$>$ has two instances in $\textit{QS}_{1}$ at positions $<$1, 2$>$ and $<$1, 3$>$.

The computation of the fuzzy utility of an $f$-item was provided in Definition \ref{fitem_utility}. It should be noted that this computation is completely different from those in previous studies \cite{lan2013mining,lan2015fuzzy}, where only the fuzzy set of the quantity (i.e., internal utility) is considered, and important external utilities are ignored. These utilities have great influence on decision making in some cases. For example, the decision-maker in a consumer electronics retail store tends to attach importance to the sequences that consist of commodities yielding a higher profit, and thus it is important to distinguish particular items generating slightly high or very high profit. Therefore, we consider not only the fuzzy set of internal utilities but also that of external utilities, and define a novel fuzzy utility calculation method for $f$-items, as shown in Definition \ref{fitem_utility}. Based on this definition, we formalize the calculation method of fuzzy utility in $f$-itemsets/$f$-sequences.

\begin{definition}
	Let \textit{QS} = $<$$\textit{QX}_{1}$, $\textit{QX}_{2}$, $\cdots$, $\textit{QX}_{n}$$>$ be a $q$-sequence, and \textit{FX} = \{(\textit{$j_{1}$}:$\textit{Region}_{k_{1}}$), (\textit{$j_{2}$}:$\textit{Region}_{k_{2}}$), $\cdots$, (\textit{$j_{m}$}:$\textit{Region}_{k_{m}}$)\} be an $f$-itemset. We assume that $\textit{FX} \sqsubseteq \textit{QX}_{j}$. Then, the fuzzy utility of \textit{FX} in the $j$-th $q$-itemset in \textit{QS} is denoted as $\textit{fu}(\textit{FX}, j, \textit{QS})$, and can be calculated as $\textit{fu}(\textit{FX}, j, \textit{QS})$ = $\sum_{\forall i \in \textit{FX}}^{}{\textit{fu}(i, j, \textit{QS})}$.
\end{definition}

For example, the $f$-itemset \textit{FX} = \{(\textit{b}:\textit{Low}) (\textit{d}:\textit{High})\} in the \textit{1st} $q$-itemset in $\textit{QS}_{1}$ can be calculated as $\textit{fu}(\textit{FX}$, $\textit{QS}_{1})$ = 2 $\times$ \$1 $\times$ 0.8 + 3 $\times$ \$4 $\times$ 1.0 = \$13.6 in TABLE \ref{table1}.  

\begin{definition}
	Let \textit{QS} = $<$$\textit{QX}_{1}$, $\textit{QX}_{2}$, $\cdots$, $\textit{QX}_{n}$$>$ be a $q$-sequence, and \textit{FS} = $<$$\textit{FX}_{1}$, $\textit{FX}_{2}$, $\cdots$, $\textit{FX}_{m}$$>$ be an $f$-sequence. We assume that $\textit{FS}$ has an instance in \textit{QS} at position $p$: $<$$k_{1}$, $k_{2}$, $\cdots$, $k_{m}$$>$. Then, the fuzzy utility of \textit{FS} in \textit{QS} at position $p$ is defined as $\textit{fu}(\textit{FS}, p, \textit{QS})$ = $\sum_{v=1}^{m}{\textit{fu}(\textit{FX}_{v}, k_{v}, \textit{QS})}$.
\end{definition}

An $f$-sequence may occur in a $q$-sequence more than once; that is, a $q$-sequence can have several instances of an $f$-sequence. Seemingly, there may be several fuzzy utilities of an $f$-sequence in a $q$-sequence. To resolve this ambiguity, we define the following. 

\begin{definition}
	Let \textit{QS} be a $q$-sequence, and \textit{FS} be an $f$-sequence. We assume that $\textit{FS}$ has $m$ instances in \textit{QS} at positions $p(\textit{FS},\textit{QS})$ = $\{p_{1}$, $p_{2}$, $\cdots$, $p_{m}\}$, where $p(\textit{FS},\textit{QS})$ represents the set of positions of all instances of \textit{FS} in \textit{QS}. We define the fuzzy utility of \textit{FS} in \textit{QS} as the maximum utility value in all $\textit{fu}(\textit{FS},\textit{QS})$ = $\max \{\textit{fu}(\textit{FS}$, $p_{v}$, $\textit{QS})| p_{v}\in p(\textit{FS},\textit{QS})$ $\forall 1 \leq v \leq m \}$. In addition, the overall fuzzy utility of an $f$-sequence \textit{FS} in a $q$-sequence database $D$ under a membership function is the sum of the fuzzy utility values in each $q$sequence, which is defined as $\textit{fu}(\textit{FS})$ = $\sum_{\forall \textit{QS}\in D}^{}{\textit{fu}(\textit{FS},\textit{QS})}$.
\end{definition}

For example, as shown in TABLE \ref{table1}, the $f$-sequence $\textit{FS}_{1}$ = $<$\{(\textit{b}:\textit{Low})\}, \{(\textit{e}:\textit{Middle})\}$>$ has two instances in $\textit{QS}_{1}$. Then, we have $\textit{fu}(\textit{FS}_{1}$, $\textit{QS}_{1})$ = $\max$\{\textit{fu}($\textit{FS}_{1}$, $<$1, 2$>$, \textit{QS}), \textit{fu}($\textit{FS}_{1}$, $<$1, 3$>$, \textit{QS})\} = $\max$\{2 $\times$ \$1 $\times$ 0.8 + 2 $\times$ \$2 $\times$ 0.4, 2 $\times$ \$1 $\times$ 0.8 + 3 $\times$ \$2 $\times$ 1.0\} = $\max$\{\$3.2, \$7.6\} = \$7.6. In addition, for another $f$-sequence  $\textit{FS}_{2}$ = $<$\{(\textit{a}:\textit{Low})\}, \{(\textit{a}:\textit{Middle})\}$>$, the fuzzy utility in the database is $\textit{fu}(\textit{FS}_{1})$ = \$0 + \$0 + \$3.0 + \$7.6 = \$10.6. Thus, it can be used for human-explainable intelligence.

\subsection{Problem Formulation}

\begin{definition}
	Given a $q$-sequence database $D$, an $f$-sequence $\textit{FS}$ is said to be a high-fuzzy-utility sequential pattern (HFUSP) if its fuzzy utility value satisfies $\textit{fu}(\textit{FS}) \geq u(D) \times \xi$, where $\xi$ is a user-specified minimum fuzzy utility threshold.
\end{definition}

\textbf{Problem statement.} Based on the above definitions, the formal statement of the problem of FUM on sequence data with linguistic meaning is defined as follows. The objective of the proposed task is to discover the complete set of HFUSPs from a $q$-sequence database, including a series of $q$-sequences and a utility table with respect to a user-defined minimum fuzzy utility threshold.

\section{Proposed Explainable PGFUM Algorithm}
\label{sec:method}

\begin{figure}[htbp]
	\centering
	\includegraphics[width=1\linewidth]{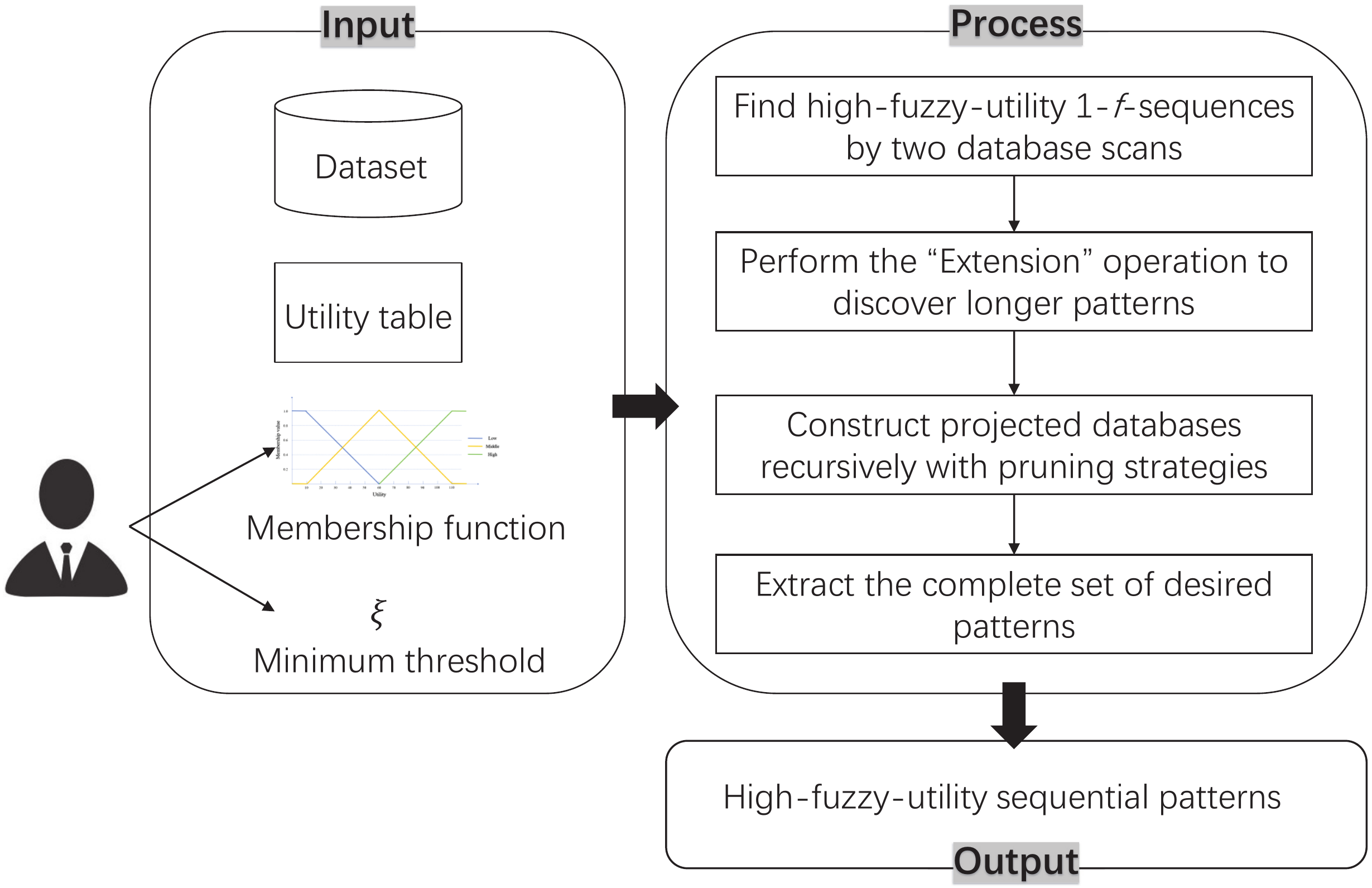}
	\vspace{-0.5cm}
	\caption{Framework of the proposed PGFUM algorithm.}
	\label{framework}
\end{figure}

According to the aforementioned definitions, we design an explainable fuzzy-theoretic algorithm named pattern growth fuzzy utility mining (PGFUM) for a highly efficient identification of the complete set of HFUSPs. Without loss of generality, we discuss the details of PGFUM in the context of a user-specified minimum utility threshold $\xi$ and a $q$-sequence database $D$ with a utility table. The whole framework of the proposed PGFUM algorithm is shown in Fig. \ref{framework}. Details are discussed as follows.

\subsection{Data Structures with Explainable Fuzziness}

As presented previously \cite{yin2012uspan,zaki2001spade,ahmed2010novel}, in the problem of conventional pattern mining, the search space can be represented as a tree structure called a prefix tree, where each node represents a candidate to be checked, except for the root, which is a null sequence. Owing to the incorporation of fuzziness, a prefix tree is not suitable for the task of FUM on sequences. Thus, based on the prefix tree, we develop a new structure called fuzzy extension tree (\textit{FE}-tree) with membership degree information. In addition, we introduce two novel data structures for storing necessary information that contributes to the rapid calculation of fuzzy utility and upper bound values. To facilitate the discussion on the three data structures, we first formulate the following concepts.

\begin{definition}
	Extension is a common addition operation on sequences, and it can be classified as $I$-Extension and $S$-Extension according to the additional positions. Let $\textit{FS}$ = $<$$\textit{FX}_{1}$, $\textit{FX}_{2}$, $\cdots$, $\textit{FX}_{m}$$>$ be an $f$-sequence. Given an $f$-item $i$, the $I$-Extension operation on $\textit{FS}$ appends $i$ to the last $f$-itemset $\textit{FX}_{m}$ to generate a new $f$-sequence $\textit{FS}'$, denoted as $<$$\textit{FS} \bigoplus i $$>$. The $S$-Extension operation of $\textit{FS}$ appends $i$ to a new empty $f$-itemset $\textit{FX}_{m+1}$, and $\textit{FX}_{m+1}$ is put behind $\textit{FX}_{m}$, resulting in a new $f$-sequence $\textit{FS}'$, denoted as $<$$\textit{FS} \bigotimes i$$>$. If the $f$-sequence $\textit{FS}'$ can be generated from $t$ by an $I$/$S$-Extension, we say that $\textit{FS}'$ can be extended from $\textit{FS}$ and is an $I$/$S$-Extension $f$-sequence of $\textit{FS}$.
\end{definition}

\begin{figure}[htbp]
	\centering
	\includegraphics[width=1\linewidth]{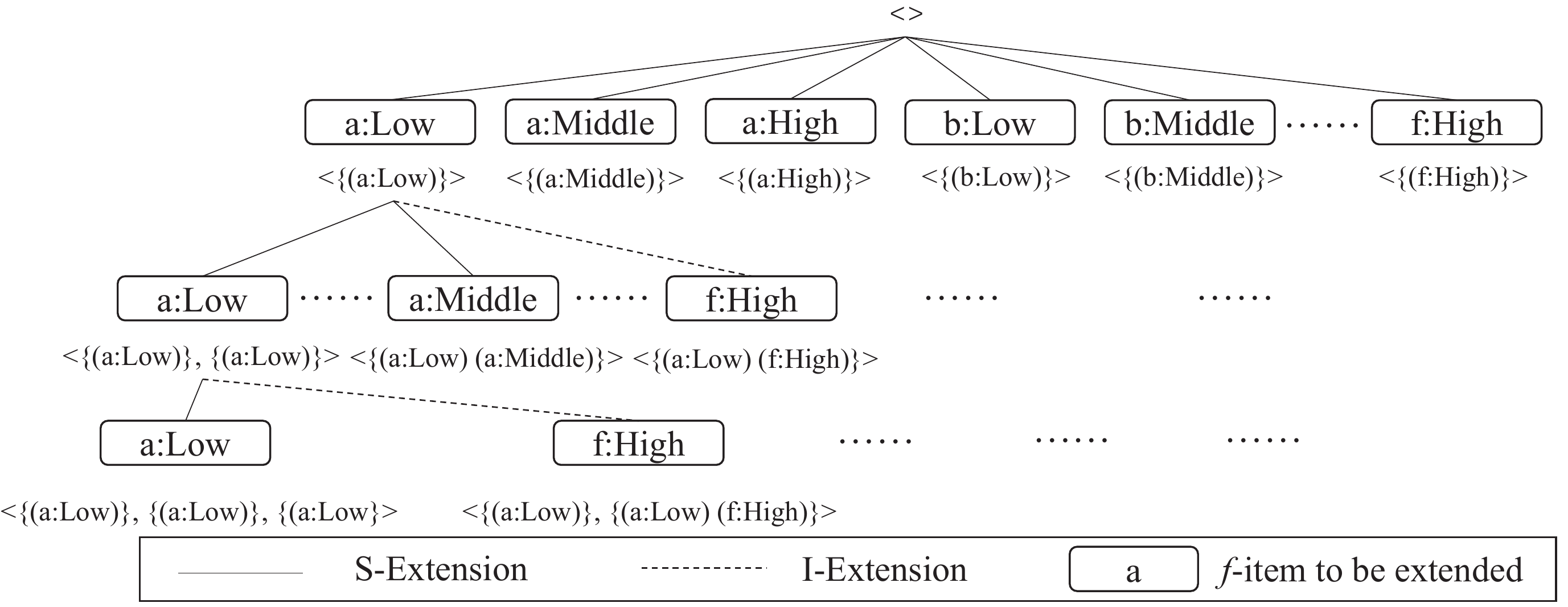}
	\vspace{-0.5cm}
	\caption{\textit{FE}-tree for the running example.}
	\label{tree}
\end{figure}

For a tree node $N$ in the \textit{FE}-tree, the children of $N$, which are arranged in alphabetical order, are $I$/$S$-Extension $f$-sequences of the $f$-sequence represented by $N$. An \textit{FE}-tree representing the search space of the running example is shown in Fig. \ref{tree}. It is worth mentioning that the \textit{FE}-tree is a conceptual structure, and in practice, the search space may differ depending on the circumstances. Generally, the complete search space may be excessively large, negatively affecting the execution efficiency of algorithms. Thus, pruning strategies have been proposed to prune needless branches of the \textit{FE}-tree and reduce the search space. The details of the proposed pruning strategies can be found in Section \ref{Pruning}. As shown in Fig. \ref{tree}, the $f$-sequences $<$\{(\textit{a}:\textit{Low})\}, \{(\textit{a}:\textit{Low})\}$>$, $<$\{(\textit{a}:\textit{Low}) (\textit{f}:\textit{Middle})\}$>$ are both extension $f$-sequences of $<$\{(\textit{a}:\textit{Low})\}$>$; therefore, the corresponding nodes are children of the node representing $<$\{(\textit{a}:\textit{Low})\}$>$. It should be noted that we omit some components in Fig. \ref{tree} for simplicity.

\begin{definition}
	\label{MFSU}
	Given a $q$-sequence \textit{QS}, the maximum fuzzy utility of an item $i$ (\textit{MFUI}) in the $j$-th $q$-itemset is defined as $\textit{MFUI}(i, j, \textit{QS})$ = $\max\{\textit{fu}(i_{1}, j, \textit{QS})$, $\textit{fu}(i_{2}, j, \textit{QS})$, $\cdots$, $\textit{fu}(i_{h}, j, \textit{QS})\}$, where $h$ is the number of regions of $i$, and $\textit{fu}(i_{m}, j, \textit{QS})$ is the fuzzy utility of $i$ in the $m$-th region. The maximum fuzzy sequence utility (\textit{MFSU}) of \textit{QS} is denoted as $\textit{MFSU}(\textit{QS})$, and is defined as the sum of the maximum fuzzy utilities of all items in \textit{QS}. \cite{lan2013mining}
\end{definition}

We consider $\textit{QS}_{1}$ in the running example in Table \ref{table1}. The item $e$ in the \textit{2nd} $q$-itemset can be calculated as $\textit{MFUI}(e, 2, \textit{QS}_{1})$ = $\max\{\$1.6, \$2.4, \$0\}$ = \$2.4 and $\textit{MFSU}(\textit{QS}_{1})$ = \$1.6 + \$12.0 + \$6.0 + \$2.4 + \$1.0 + \$12.0 + \$6.0 = \$41.0.

\begin{definition}
	\label{MRFU}
	We assume that there are an $f$-sequence \textit{FS} and a $q$-sequence \textit{QS}, and that one of the instances of \textit{FS} in \textit{QS} is at position $p$: $<$$k_1$, $k_2$, $\cdots$, $k_m$$>$. Then, we say that the extension position of this instance is $k_m$, and the last item in \textit{FS} is called the extension item. Furthermore, we assume that there are $n$ extension positions of \textit{FS} in \textit{QS}: \{$p_1$, $p_2$, $\cdots$, $p_n$\}. Then, the fuzzy utility of \textit{FS} in \textit{QS} at extension position $p_i$, denote as $\textit{fu}(\textit{FS}$, $p_i$, $\textit{QS})$, is defined as $\textit{fu}(\textit{FS}$, $p_i$, $\textit{QS})$ = $\max\{\textit{fu}(\textit{FS}$, $<$$j_1$, $j_2$, $\cdots$, $p_i$$>$, $s)|<$$j_1$, $j_2$, $\cdots$, $p_i$$> \in P(\textit{FS},\textit{QS})\}$, where $<$$j_1$, $j_2$, $\cdots$, $p_i$$>$ is the position of an instance with extension position $p_i$. The maximum remaining fuzzy utility (\textit{MRFU}) of \textit{FS} at each extension position $p_i$, denoted as $\textit{MRFU}(\textit{FS}$, $p_i$, $\textit{QS})$, is defined as $\textit{MRFU}(\textit{FS}$, $p_i$, $\textit{QS})$ = $\sum_{i'\subseteq \textit{QS} \land i'\succ p_i}^{}\textit{MFUI}(i, j', \textit{QS})$, where $i'\succ p_i$ indicates that the position of $i'$ is after the extension position $p_i$, and $j \leq j'$.
\end{definition}

\begin{figure}[htbp]
	\centering
	\includegraphics[width=1\linewidth]{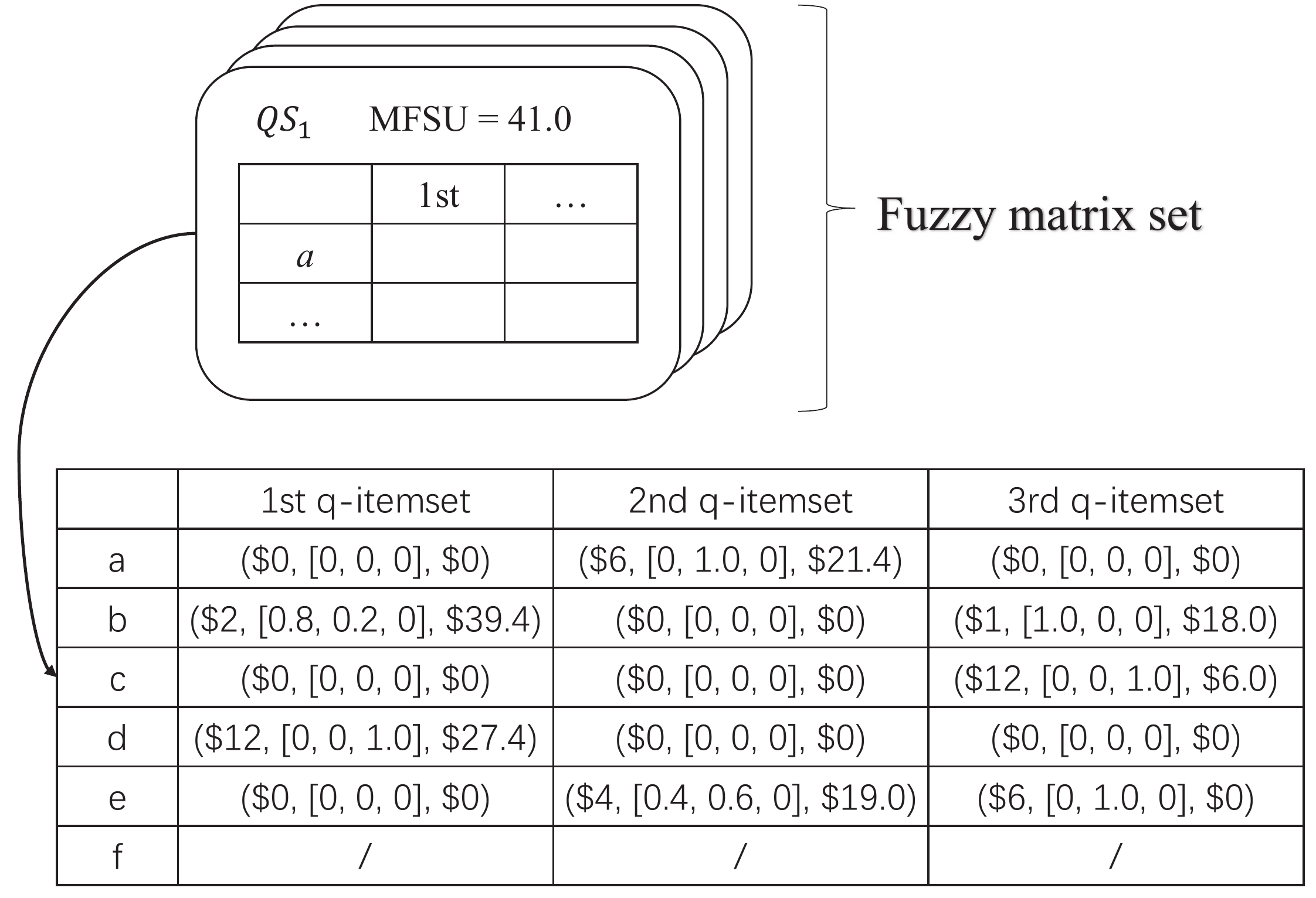}
	\vspace{-0.5cm}
	\caption{$F$-matrix set for the running example.}
	\label{matrix}
\end{figure}

In the example shown in Table \ref{table1}, the $f$-sequence \textit{FS}: $<$\{(\textit{a}:\textit{Middle})\}, \{(\textit{e}:\textit{Middle})\}$>$ has two instances in $\textit{QS}_{2}$ at positions $<$1, 3$>$ and $<$2, 3$>$, the extension positions of which are both 3. Then, we have $\textit{fu}(\textit{FS}, 3, \textit{QS}_{2})$ = $\max\{\$6.4,\ \$5.2\}$ = \$6.4 and $\textit{MRFU}(\textit{FS}, 3, \textit{QS}_{2})$ = $\textit{MFUI}(\textit{d}, 4, {QS}_{2})$ = \$20. In another case, $\textit{MRFU}(\textit{FS}, 2, \textit{QS}_{3})$ = $\textit{MFUI}(\textit{a}, 3, \textit{QS}_{3})$ + $\textit{MFUI}(\textit{d}, 3, \textit{QS}_{3})$ +$\textit{MFUI}(\textit{e}, 3, \textit{QS}_{3})$ = \$2.4 + \$12 + \$1.6 = \$16.0. Every input $q$-sequence is stored in the primary storage in the form of a fuzzy matrix ($f$-matrix), all of which together with their identifiers and \textit{MFSU} values compose a data structure termed $f$-matrix set. The $f$-matrix set for the running example is shown in Fig. \ref{matrix}. The $f$-matrix of $\textit{QS}_{1}$ under the membership function is shown in detail in Fig. \ref{membership_function}, and the $f$-matrices of the other $q$-sequences can be obtained in the same manner. As can be observed in Fig. \ref{matrix}, each element indexed by an item and a $q$-itemset in the $f$-matrix is a tuple with three fields: 1) the first field indicates the utility of the $q$-item, 2) the second is a list consisting of a series of membership values, and 3) the third is the sum of the \textit{MFUI} values of the items after the $q$-item. For an $f$-matrix, the element is empty when the item does not appear in the $q$-sequence, and all terms are set to 0 if the item is contained in the $q$-itemset. 

For example, we consider the entry (\textit{a}, \textit{2nd} $q$-itemset) in the given $f$-matrix of $\textit{QS}_{1}$. The first term $u(\textit{a}, 2, \textit{QS}_{1})$ can be calculated as 3 $\times$ \$2 = \$6, and the last is equal to $\sum_{i'\subseteq \textit{QS} \land i'\succ \textit{a}}^{}\textit{MFUI}(i, 2, \textit{QS}_{3})$ = \$2.4 + \$1 + \$12 + \$6 = \$21.4, where \textit{a} refers, in particular, to the item in the \textit{2nd} $q$-itemset. Based on the membership function with the three regions in Fig. \ref{membership_function}, the calculation results of the three membership values are 0, 1.0, and 0. The terms of the entry (\textit{a}, \textit{1st} $q$-itemset) are all set to 0 because \textit{a} is not contained in this $q$-itemset. In addition, we can observe that the item \textit{f} does not appear in $\textit{QS}_{1}$; thus, the corresponding row is null.

\begin{figure}[htbp]
	\centering
	\includegraphics[width=1\linewidth]{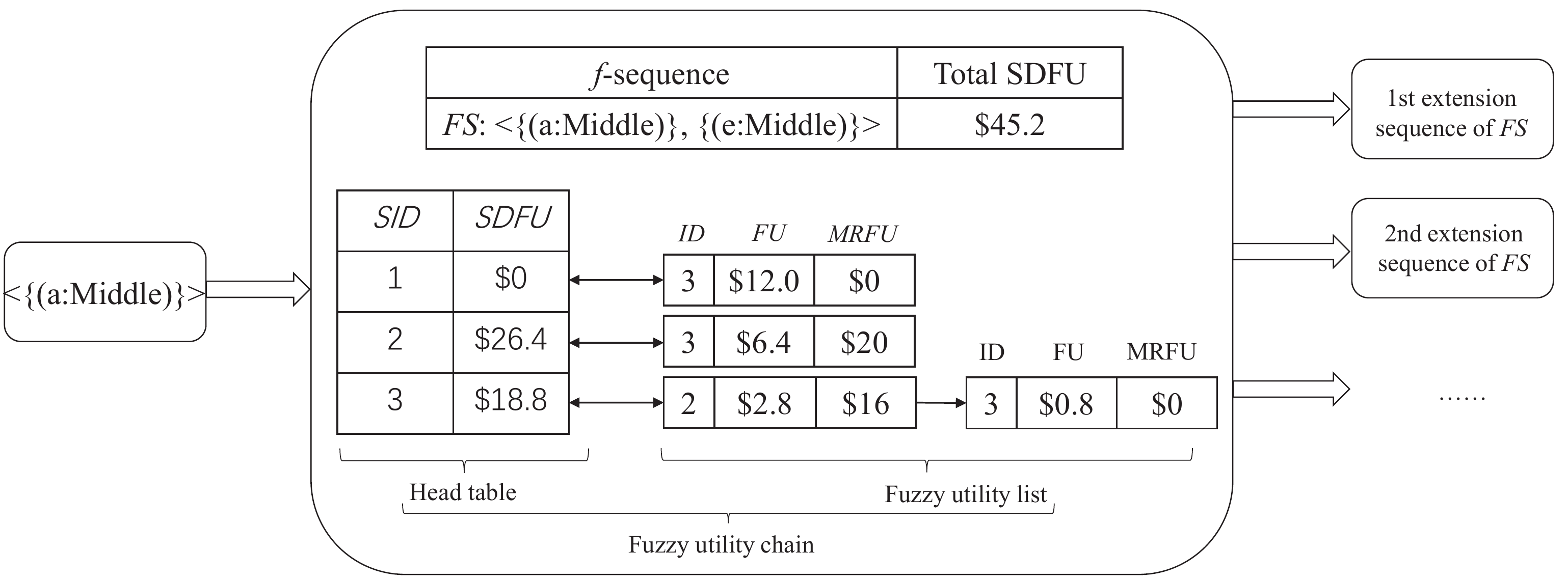}
	\vspace{-0.5cm}
	\caption{Fuzzy utility chain the $f$-sequence \textit{FS}: $<$\{(\textit{a}:\textit{Middle}) (\textit{e}:\textit{Middle})\}$>$.}
	\label{chain}
\end{figure}

According to the aforementioned calculation method for fuzzy utility, a simple and intuitive approach to verify whether a candidate is an HFUSP is to scan the entire database. However, a brute-force search strategy incurs high computational costs, particularly in large databases. In view of this, we design a compact data structure termed fuzzy utility chain to be the projected database for storing the necessary information of each candidate in the search process. The projected database of the $f$-sequence \textit{FS}: $<$\{(\textit{a}:\textit{Middle}) (\textit{e}:\textit{Middle})\}$>$ is shown as an example in Fig. \ref{chain}. The fuzzy utility chain consists of a head table and multiple fuzzy utility lists. Each fuzzy list can be indexed by an entry in the head table, which includes two fields: \textit{SID} and \textit{SDFU}. The former is the identifier of the $q$-sequence containing \textit{FS}, whereas the latter is an upper bound value of \textit{FS} in this $q$-sequence. We discuss the details on \textit{SDFU} in the next subsection. We now turn our attention to the utility list consisting of several fuzzy utility elements, each of which corresponds to an extension position of \textit{FS} in the current $q$-sequence. It is assumed that \textit{FS} has $m$ extension positions $\textit{EP}$ = $\{p_1$, $p_2$, $\cdots$, $p_m\}$ in the $q$-sequence \textit{QS}. The fuzzy utility element in the utility list of \textit{QS} contains the following fields: 1) the field \textit{ID} is the $i$-th extension position $p_i$, 2) the field \textit{FU} indicates the fuzzy utility of \textit{FS} at the $i$-th extension position $p_i$ (i.e., $\textit{fu}(\textit{FS}$, $p_i$, $\textit{QS})$), and 3) the field \textit{MRFU} is the \textit{MRFU} value of \textit{FS} at the $i$-th extension position $p_i$ (i.e., $\textit{MRFU}(\textit{FS}$, $p_i$, $\textit{QS})$). 

To calculate the fuzzy utility of an $f$-sequence \textit{FS}, scanning the small-scale projected database is sufficient. This is because all the necessary information of $q$-sequences containing \textit{FS} is not only exact but also complete, and unpromising $q$-sequences not containing \textit{FS} do not occur in the projected database. We note that the projected database of the ($l$+1)-sequence can be constructed using the projected database of its prefix, which is an $l$-sequence, rather than a complete scan of the original database. The proposed PUFUM algorithm adopts a projection strategy that can limit the scope of database scans by recursively constructing the projected database of the current candidate and its extension $f$-sequences. As shown in Fig. \ref{chain}, the projected database of an $f$-sequence $<$\{(\textit{a}: \textit{Middle}) (\textit{e}: \textit{Middle})\}$>$ is constructed from that of its prefix $<$\{(\textit{a}: \textit{Middle})\}$>$. Analogously, a series of projected databases of its extension $f$-sequences can be constructed in the same manner if the pruning strategy conditions in Section \ref{Pruning} are satisfied. This divide-and-conquer technique greatly reduces the high computational costs of scanning and improves efficiency.

\subsection{Upper Bounds of Explainable Fuzzy Utility}

As can be seen in Fig. \ref{tree}, there is a combinatorial explosion of the search space in FUM on sequences; that is, the \textit{FE}-tree structure is excessively large. In particular, a large number of types of $f$-items appear. Other tasks, such as HUSPM \cite{gan2020survey}, FIM \cite{han2004mining}, and SPM \cite{gan2019survey}, also involve the same challenge. To handle this issue, several upper bounds have been proposed to facilitate the pruning of the tree structure in HUSPM. For example, \textit{SWU} \cite{ahmed2010novel}, \textit{MEU} \cite{lin2017high}, \textit{CRoM} \cite{alkan2015crom}, and \textit{PEU} \cite{wang2016efficiently}. Following the concept of upper bounds on maximum utility in HUSPM algorithms, the first and only high fuzzy-sequence-utility upper bound (\textit{HFSUUB}) was used in PFUS \cite{lan2013mining} for FUM on sequences. To improve efficiency, we further develop two novel upper bounds for sequence descendant fuzzy utility (\textit{SDFU}) and extension item fuzzy utility (\textit{EIFU}), which are considerably tighter than \textit{HFSUUB} on explainable fuzzy utility, and they are adopted in three novel pruning strategies to substantially reduce the search space and thus expedite the mining process.

\textit{\textbf{HFSUUB}} is a relatively loose upper bound, but it is quite effective in the early stage of the mining process because it is simple and easy to calculate. Based on Definition \ref{MFSU}, we state the \textit{HFSUUB} upper bound as follows: The \textit{HFSUUB} value of an $f$-sequence \textit{FS} is the sum of the \textit{MFSU} values of all $q$-sequences containing \textit{FS} in a $q$-sequence $D$:
\begin{equation}
	\textit{HFSUUB}(\textit{FS}) = \sum_{\textit{QS} \in D \land \textit{FS} \subseteq \textit{QS}}^{}\textit{MFSU(QS)}
\end{equation}

We consider the $f$-sequence \textit{FS} = $<$\{(\textit{e}:\textit{Low})\}$>$ in Fig. \ref{table1}; evidently, it is contained in $\textit{QS}_1$, $\textit{QS}_2$, and $\textit{QS}_3$. Therefore, we have \textit{HFSUUB}(\textit{FS}) = \textit{MFSU($\textit{QS}_1$)} + \textit{MFSU($\textit{QS}_2$)} + \textit{MFSU($\textit{QS}_3$)} = \$41.0 + \$57.6 + \$26.0 = \$124.6.

In addition, we develop two novel and tighter upper bounds \textit{SDFU} and \textit{EIFU}, which are adopted in the recursive mining process.

\textbf{The \textit{SDFU} value of a $f$-sequence} is an upper bound of the fuzzy utility of its descendants in the \textit{FE}-tree; we prove this in the next subsection. We assume that there exist an $f$-sequence \textit{FS} and a $q$-sequence \textit{QS}, and \textit{FS} has an instance in \textit{QS} at extension position $p$. Based on Definition \ref{MRFU}, we define the \textit{SDFU} upper bound of \textit{FS} in \textit{QS} at $p$ as 
\begin{equation}
\textit{SDFU}(\textit{FS},p,\textit{QS})\ =\ \textit{fu}(\textit{FS},p,\textit{QS})\ +\ \textit{MRFU}(\textit{FS},p,\textit{QS}).
\end{equation}
If the inequality $\textit{MRFU}(\textit{FS}$, $p$, $\textit{QS})>0$ holds; otherwise, $\textit{SDFU}(\textit{FS}$, $p$, $\textit{QS})$ = 0. We assume that \textit{FS} has several instances in \textit{QS} at $m$ extension positions $\textit{EP}$ = $\{p_1$, $p_2$, $\cdots$, $p_m\}$. The \textit{SDFU} value of \textit{FS} in \textit{QS} is the maximum value of \textit{SDFU} at all extension positions. That is, 
\begin{equation}
\textit{SDFU}(\textit{FS},\textit{QS}) = \max\{ \textit{SDFU}(\textit{FS}, p_{k},\textit{QS})| \forall 1 \leq k \leq m\}.
\end{equation}

Moreover, the \textit{SDFU} value of \textit{FS} in a $q$-sequence $D$ can be defined as:
\begin{equation}
\textit{SDFU}(\textit{FS}) = \sum_{\textit{QS} \in D \land \textit{FS} \subseteq \textit{QS}}^{}\textit{SDFU(\textit{FS},\textit{QS})}.
\end{equation}

For instance, we consider an $f$-sequence \textit{FS} = $<$\{(\textit{a}: \textit{Middle})\}, \{(\textit{e}: \textit{Middle})\}$>$. Then, we have $\textit{SDFU}(\textit{FS}, 2, \textit{QS}_3)$ = \$2.8 + \$16 = \$18.8, $\textit{SDFU}(\textit{FS}$, $\textit{QS}_3)$ = $\max$\{\$18.8, \$0.8\} = \$18.8, and $\textit{SDFU}(\textit{FS})$ = \$0 + \$26.4 + \$18.8 = \$45.2.

\textbf{The \textit{EIFU} value of an $f$-sequence} is an upper bound of its fuzzy utility and its descendants in the \textit{FE}-tree. The proof can also be found in the next subsection. It is assumed that the $f$-sequence \textit{FS} is an extension $f$-sequence of the $f$-sequence $\alpha$, and we define the \textit{EIFU} of \textit{FS} in a $q$-sequence \textit{QS} as 
\begin{equation}
	\textit{EIFU}(\textit{FS},\textit{QS})\begin{cases}
\textit{SDFU}(\alpha,\textit{QS}) & \alpha \subseteq \textit{QS} \land \textit{FS} \subseteq \textit{QS}\\
0 & otherwise
\end{cases}
\end{equation}

Similarly, the \textit{EIFU} value of \textit{FS} in a $q$-sequence $D$ can be defined as:
\begin{equation}
	\textit{EIFU}(\textit{FS}) = \sum_{\textit{QS} \in D \land \textit{FS} \subseteq \textit{QS}}^{}\textit{EIFU(\textit{FS},\textit{QS})}.
\end{equation}

For example, we consider the $f$-sequence $\alpha$ = $<$\{(\textit{a}: \textit{Middle})\},  \{(\textit{e}: \textit{Middle})\}$>$, and let $\textit{FS}$ be an $S$-Extension $f$-sequence and $\textit{FS}$ = $<$\{(\textit{a}: \textit{Middle})\}, \{(\textit{e}: \textit{Middle})\}, \{(\textit{d}: \textit{High})\}$>$. We have $\textit{EIFU}(\textit{FS}$, $\textit{QS}_3)$ = $\textit{SDFU}(\textit{FS}$, $\textit{QS}_3)$ \$18.8 because $\textit{FS}$ and $\alpha$ are both contained in $\textit{QS}_3$, and  $\textit{EIFU}(\textit{FS})$ = \$26.4 + \$18.8 = \$45.2.

\subsection{Pruning Strategies with Explainable Upper Bounds}
\label{Pruning}

Herein, we propose three new efficient pruning strategies. It is a key contribution of this study that these strategies can prune the search space represented by the \textit{FE}-tree structure, thus expediating the mining process. Evidently, if more candidates are eliminated by pruning, the desired patterns can be discovered more efficiently. Therefore, we adopt the three pruning strategies in different mining strategies to reduce unnecessary searches.

\begin{theorem}
	\label{HFSUUB}
	Given a $q$-sequence database $D$ and two $f$-sequence $\textit{FS}$ and $\textit{FS}'$, we assume that $\textit{FS}$ is the same as $\textit{FS}'$, or the node representing $\textit{FS}$ is the descendant of the node representing $\textit{FS}'$ (i.e., $\textit{FS}'$ is the prefix of $\textit{FS}$). Then, 
	\begin{equation}
	\textit{fu}(\textit{FS}) \leq \textit{HFSUUB}(\textit{FS}').
	\end{equation}
\end{theorem}

\begin{proof}
	As $\textit{FS}$ is the same as $\textit{FS}'$, or $\textit{FS}'$ is the prefix of $\textit{FS}$, a $q$-sequence in the $q$-sequence database $D$ contains $\textit{FS}$, and therefore it must contain $\textit{FS}'$. By Definition \ref{MFSU}, $\textit{fu}(\textit{FS},\textit{QS}) \leq \textit{MFSU}(\textit{QS})$. Therefore, we have $\textit{fu}(\textit{FS}) \leq \sum_{\textit{QS} \in D \land t \subseteq \textit{QS}}^{}{\textit{MFSU}(\textit{QS})} \leq \sum_{\textit{QS} \in D \land \textit{FS}' \subseteq \textit{QS}}^{}{\textit{MFSU}(\textit{QS})} \leq \textit{HFSUUB}(\textit{QS}).$
\end{proof}

Based on the \textit{HFSUUB} upper bound and Theorem \ref{HFSUUB}, we describe the first pruning strategy \textbf{Pre-pruning 1-\textit{f}-sequences} (\textbf{PPO}) as follows. Given a 1-$f$-sequence $\textit{FS}$ represented by a node $N$ in the \textit{FE}-tree, a minimum fuzzy utility threshold $\xi$, and a $q$-sequence database $D$, if $\textit{FS}$ satisfies $\textit{HFSUUB(\textit{FS})}$ $< \xi \times u(D)$, then PGFUM can stop exploring node $N$. This implies that $\textit{FS}$ and the $f$-sequences represented by the descendants of $N$ can be regarded as unpromising candidates with no possibility to be an HFUSP because of the downward closure property of \textit{HFSUUB}, as proven above.

\begin{theorem}
	\label{t_SDFU}
	Given a $q$-sequence database $D$ and two $f$-sequences $\textit{FS}$ and $\textit{FS}'$, we assume that the node representing $\textit{FS}'$ is the descendant of the node representing $\textit{FS}$ (i.e., $\textit{FS}'$ is the prefix of $\textit{FS}$). Then, 
	\begin{equation}
	\textit{fu}(\textit{FS}) \leq \textit{SDFU}(\textit{FS}').
	\end{equation}
\end{theorem}

\begin{proof}
	We denote $\textit{FS}$ as $\textit{FS}' \bullet \textit{FS}''$, where $|\textit{FS}''| > 0$ and $\bullet$ is the usual concatenation operation on sequences, as $\textit{FS}'$ is a prefix $f$-sequence of $\textit{FS}$. It is assumed that $t$ is contained in a $q$-sequence \textit{QS}, and thus $\textit{FS}''$ is contained in \textit{QS}. The fuzzy utility of $t$ in \textit{QS} can be divided into two parts as $\textit{fu}(\textit{FS}$, $\textit{QS})$ = $\textit{fu}(\textit{FS}'$, $p$, $\textit{QS})$ + $\textit{fu}_{\textit{conditional}}(\textit{FS}'')$, where $\textit{fu}(\textit{FS}'$, $p$, $\textit{QS})$ denotes the fuzzy utility of an instance of $\textit{FS}'$ at extension position $p$ in \textit{QS}, and $\textit{fu}_{\textit{conditional}}(\textit{FS}'')$ is the fuzzy utility of an instance of $\textit{FS}''$ in $s$ under the condition that the position of the first item of $\textit{FS}''$ is after $p$. Clearly, $\textit{fu}_{\textit{conditional}}(\textit{FS}'')$ $< \textit{MRFU}(\textit{FS}'$, $p$, $\textit{QS})$; then, we have
	\begin{align*}
	\textit{fu}(\textit{FS},\textit{QS}) &= \textit{fu}(\textit{FS}', p, \textit{QS}) + \textit{fu}_{\textit{conditional}}(\textit{FS}'')  \\
	&\leq \textit{fu}(\textit{FS}', p, \textit{QS}) + \textit{MRFU}(\textit{FS}', p, \textit{QS}),  \\
	&\leq max\{\textit{fu}(\textit{FS}', p, \textit{QS}) + \textit{MRFU}(\textit{FS}', p, \textit{QS})\}  \\
	&\leq \textit{SDFU}(\textit{FS}',\textit{QS}).
	\end{align*}
	The $q$-sequence containing $\textit{FS}$ must contain $\textit{FS}'$ because $\textit{FS}'$ $\sqsubseteq t$; thus, we have $\textit{fu}(\textit{FS})$ = $\sum_{\textit{QS} \in D \land t \sqsubseteq \textit{QS}}^{}\textit{fu}(\textit{FS},\textit{QS})$ $\leq$ $\sum_{\textit{QS} \in D \land \textit{FS}' \sqsubseteq \textit{QS}}^{}{}\textit{SDFU}(\textit{FS}',\textit{QS})$ = $\textit{SDFU}(\textit{FS}')$ in the $q$-sequence database $D$.
\end{proof}

Adopting the \textit{SDFU} upper bound, we design the second pruning strategy 
\textbf{eliminating unpromising descendants} (\textbf{EUD}), which can be stated as follows: Given an $f$-sequence $\textit{FS}$ represented by a node $N$ in the \textit{FE}-tree, a minimum fuzzy utility threshold $\xi$, and a $q$-sequence database $D$, PGFUM can eliminate the $f$-sequences represented by the descendants of $N$ if $\textit{SDFU(\textit{FS})}$ $< \xi \times u(D)$. By Theorem \ref{t_SDFU}, we have proven the downward closure property of \textit{SDFU}. Thus, \textit{EUD} is also a safe strategy and will not miss any HFUSP.

\begin{theorem}
	\label{t_EIFU}
	Given a $q$-sequence database $D$ and two $f$-sequences $\textit{FS}$ and $\textit{FS}'$, we assume that $\textit{FS}$ is the same as $\textit{FS}'$, or the node representing $\textit{FS}$ is the descendant of the node representing $\textit{FS}'$ (i.e., $\textit{FS}'$ is the prefix of $\textit{FS}$). Then, 
	\begin{equation}
	\textit{fu}(\textit{FS}) \leq \textit{EIFU}(\textit{FS}').
	\end{equation}
\end{theorem}

\begin{proof}
	It is assumed that $\textit{FS}'$ is the extension $f$-sequence of $\alpha$; thus, $\alpha$ is also a prefix of $\textit{FS}$ because $\textit{FS}'$ is a prefix of $\textit{FS}$ or $\textit{FS}'$ = $\textit{FS}$.  By Theorem \ref{t_SDFU}, we have $\textit{fu}(\textit{FS}$, $\textit{QS}) \leq \textit{SDFU}(\alpha$, \textit{QS}) in a $q$-sequence \textit{QS}. By the definition of \textit{RSU}, we have that $\textit{EIFU}(\textit{FS}'$, \textit{QS}) = \textit{SDFU}($\alpha,\textit{QS})$ if \textit{QS} contains both $\textit{FS}'$ and $\alpha$. In this case, we can obtain $\textit{fu}(\textit{FS}, \textit{QS}) \leq \textit{EIFU}(\textit{FS}'$, \textit{QS}). Moreover, if \textit{QS} does not contain $\textit{FS}'$, \textit{QS} must not contain $t$ by Theorem \ref{t_EIFU}, and we obtain $\textit{fu}(\textit{FS}$, \textit{QS}) = $\textit{EIFU}(\textit{FS}'$, \textit{QS}) = 0. In conclusion, we have $\textit{fu}(\textit{FS}, \textit{QS})$  $\leq \textit{EIFU}(\textit{FS}', \textit{QS})$ in a $q$-sequence \textit{QS}. Thus, $\textit{fu}(\textit{FS})$  $\leq \textit{EIFU}(\textit{FS}')$ in the database $D$.
\end{proof}

As can be seen from Theorem \ref{t_EIFU}, the \textit{EIFU} upper bound also has global monotonicity. To further improve efficiency, we develop the third pruning strategy termed  \textbf{ pruning extension \textit{f}- sequence } (\textbf{PES}): Given an $f$-sequence $t$ represented by a node $N$ in the \textit{FE}-tree, a minimum fuzzy utility threshold $\xi$, and a $q$-sequence database $D$, PGFUM can prune $N$ and its descendants if $\textit{EIFU(\textit{FS})}$ $< \xi \times u(D)$. 

It should be noticed that in the recursive mining process, EUD is a depth-first pruning strategy, whereas PES is a width-first strategy. It is noted that for an $f$-sequence $\textit{FS}$ represented by node $N$ in the \textit{FE}-tree, the \textit{HFSUUB} and \textit{EIFU} values are the upper bounds of the fuzzy utility values of $\textit{FS}$ and the $f$-sequences represented by the descendants of $N$. In contrast to \textit{HFSUUB} and \textit{EIFU}, \textit{SDFU} is a tight upper bound of the fuzzy utility values of the $f$-sequences represented by the descendants of $N$, and the fuzzy utility of $\textit{FS}$ may be larger than its \textit{SDFU} value.

\subsection{Explainable PGFUM Algorithm}

Based on the data storing structures $f$-matrix set and fuzzy utility chain, the three upper bounds, and the three designed pruning strategies, the proposed explainable fuzzy-theoretic PGFUM algorithm is described as follows. To facilitate the presentation, we provide two pieces of pseudocode of PGFUM in Algorithms \ref{alg:PGFUM} and \ref{alg:RM}, which describe the main and the recursive mining process, respectively. 

\begin{algorithm}[htbp]
	\caption{ MainPGFUM Algorithm}
	\label{alg:PGFUM}
	\begin{algorithmic}[1]
		\REQUIRE 
		$D$: a $q$-sequence database; \textit{UT}: a table containing the external utility of each item; $f$: a membership function; $\xi$: a minimum fuzzy utility threshold.
		\ENSURE 
		a complete set of \textit{HFUSP}. 
		
		\STATE scan the original database $D$ to: \\
		(a) convert the utility of each $q$-item in each $q$-sequence to a fuzzy set by the given membership function $f$;\\
		(b) calculate the maximum fuzzy sequence utility (\textit{MFSU}) value of each $q$-sequence;\\
		(c) construct the $f$-matrix set of $D$.
		\STATE second scan the original database $D$ to: \\
		(a) calculate the utility values of all 1-$f$-sequences; \\
		(b) calculate the \textit{HFSUUB} values of all 1-$f$-sequences;\\
		(c) construct projected databases of all 1-$f$-sequences;
		
		\FOR {$\textit{FS} \in $ 1-$f$-sequences }
		\IF{$\textit{HFSUUB}(\textit{FS}) \geq u(D)\times\xi$ (\textbf{The PPO strategy})}
		\IF{$t.\textit{fu} \ge u(D)\times\xi$}
		\STATE update $\textit{HFUSP}\ \leftarrow\ \textit{HFUSP}\ \cup\ \textit{FS}$;
		\ENDIF
		\IF{$\textit{SDFU}(\textit{FS}) \ge u(D)\times\xi$ (\textbf{The EUD strategy})}
		\STATE call \textit{RecursiveMining($t,\textit{ProjectedDatabase}_\textit{FS}$)};
		\ENDIF
		\ENDIF
		\ENDFOR
		
		\RETURN \textit{HFUSP}
	\end{algorithmic}	
\end{algorithm}

As can be seen in Algorithm \ref{alg:PGFUM}, the MainPGFUM procedure takes a $q$-sequence database $D$, a utility table \textit{UT}, a membership function $f$, and a minimum fuzzy utility threshold $\xi$ as input and outputs the set of \textit{HFUSPs}. Initially, the utility values are calculated using the external utilities in the utility table \textit{UT} and the internal utilities in the initial $q$-sequence database $D$. PGFUM transforms utility values into fuzzy sets, calculates \textit{MFSU} values, and constructs the $f$-matrix set in the first scan of $D$ (line 1). Subsequently, PGFUM generates the root node with an empty sequence and begins mining with the second scan of $D$; utilities, \textit{HFSUUB} values, and projected databases of all 1-$f$-sequences can be obtained after scanning (line 2). For each node represented by a 1-$f$-sequence (lines 3--12), PGFUM adopts a PPO pruning strategy with the \textit{HFSUUB} upper bound to determine whether to continue exploring (line 4). In addition, the 1-$f$-sequence will be added to \textit{HFUSP} as a HFUSP if its fuzzy utility is larger than $u(D)\times\xi$ (lines 5--7). Finally, PGFUM checks whether the descendants of the current 1-$f$-sequence cannot be HFUSPs by the EDU pruning strategy (lines 8--10). If so, it backtracks to the root; otherwise, it calls the RecursiveMining procedure to recursively extract HFUSPs with the prefix of the current 1-$f$-sequence.

\begin{algorithm}[htbp]
	\caption{RecursiveMining}
	\label{alg:RM}
	\begin{algorithmic}[1]
		\REQUIRE 
		$\textit{FS}$: an $f$-sequence as a prefix; $\textit{ProjectedDatabase}_\textit{FS}$: the projected database of $\textit{FS}$.
		\ENSURE 
		a set of \textit{HFUSP}.
		
		\FOR {each fuzzy utility chain \textit{uc} in $\textit{ProjectedDatabase}_\textit{FS}$}
		\STATE get the $f$-matrix \textit{fm} corresponding to \textit{uc};
		\STATE find $I$-Extension $f$-items of $\textit{FS}$ in \textit{fm} and add them to \textit{ilist};
		\STATE find $S$-Extension $f$-items of $\textit{FS}$ in \textit{fm} and add them to \textit{slist};
		\ENDFOR
		
		\FOR {each $f$-item $i \in \textit{ilist}$ }
		\STATE $\textit{FS}'$ $\leftarrow$ $<$$\textit{FS} \oplus i$$>$; 
		\IF{$\textit{EIFU}(\textit{FS}')<u(D)\times\xi$ (\textbf{The PES strategy})}
		\STATE remove $i$ from \textit{ilist};
		\ENDIF
		\STATE build projected database of $\textit{FS}'$ $\textit{ProjectedDatabase}_{\textit{FS}'}$; 
		\STATE put $\textit{FS}'$ into \textit{seqlist};
		\ENDFOR
		
		\FOR {each $f$-item $i \in \textit{slist}$ }
		\STATE $\textit{FS}'$ $\leftarrow$ $<$$\textit{FS} \otimes s$$>$; 
		\IF{$\textit{EIFU}(\textit{FS}')<u(D)\times\xi$ (\textbf{The PES strategy})}
		\STATE remove $i$ from \textit{slist};
		\ENDIF
		\STATE build projected database of $\textit{FS}'$ $\textit{ProjectedDatabase}_{\textit{FS}'}$; 
		\STATE put $\textit{FS}'$ into \textit{seqlist};
		\ENDFOR
		
		\FOR {each $f$-sequence $\textit{\textit{FS}'} \in \textit{seqlist}$ }
		\IF{$\textit{FS}'.\textit{fu} \geq u(D)\times\xi$}
		\STATE update $\textit{HFUSP}\ \leftarrow\ \textit{HFUSP}\ \cup\ \textit{FS}'$;
		\ENDIF
		\IF{$\textit{SDFU}(\textit{FS}') \geq u(D) \times \xi$ (\textbf{The EUD strategy})}
		\STATE call \textit{RecursiveMining}($\textit{FS}'$, \textit{ProjectedDatabase}$_{\textit{FS}'}$);
		\ENDIF
		\ENDFOR
	\end{algorithmic}	
\end{algorithm}

The recursive search procedure is presented in Algorithm \ref{alg:RM}, which takes an $f$-sequence \textit{FS} and its projected database as input. Enumerating the $f$-sequences in ascending alphabetical order, PGFUM performs a depth-first search to extract all HFUSPs. For each fuzzy utility chain, PGFUM initializes two sets \textit{ilist} and \textit{slist}, which are used to store $I$-Extension and $S$-Extension $f$items, respectively, to the empty sequence. Then, the corresponding $f$-matrix is scanned to obtain the $f$-items for $I$-Extension and $S$-Extension (lines 1--5). For an $f$-item $i$ in \textit{ilist}, PGFUM generates $\textit{FS}'$ from \textit{FS} by an $I$-Extension operation with $i$ (line 7). Furthermore, $\textit{FS}'$ is discarded if its \textit{EIFU} upper bound, which is calculated as fuzzy utility chains are scanned, is less than the minimum utility value (lines 8--10). After adopting the PES pruning strategy, PGFUM constructs the projected database of $\textit{FS}'$ and puts $\textit{FS}'$ into the set \textit{seqlist} (lines 11--12). The $f$-items in \textit{slist} can be processed in a similar manner (lines 14--21). For each $f$-sequence $\textit{FS}'$ in \textit{seqlist}, PGFUM checks whether $\textit{FS}'$ is a HFUSP and updates \textit{HFUSP} (lines 23--25). Finally, using the EDU pruning strategy, PGFUM recursively calls the RecursiveMining procedure to continue mining the complete set of HFUSPs with respect to the prefix $\textit{FS}'$ (lines 26--28).

\section{Experiments} 
\label{sec:experiments}

\begin{figure}[htbp]
	\centering
	\includegraphics[scale=0.25]{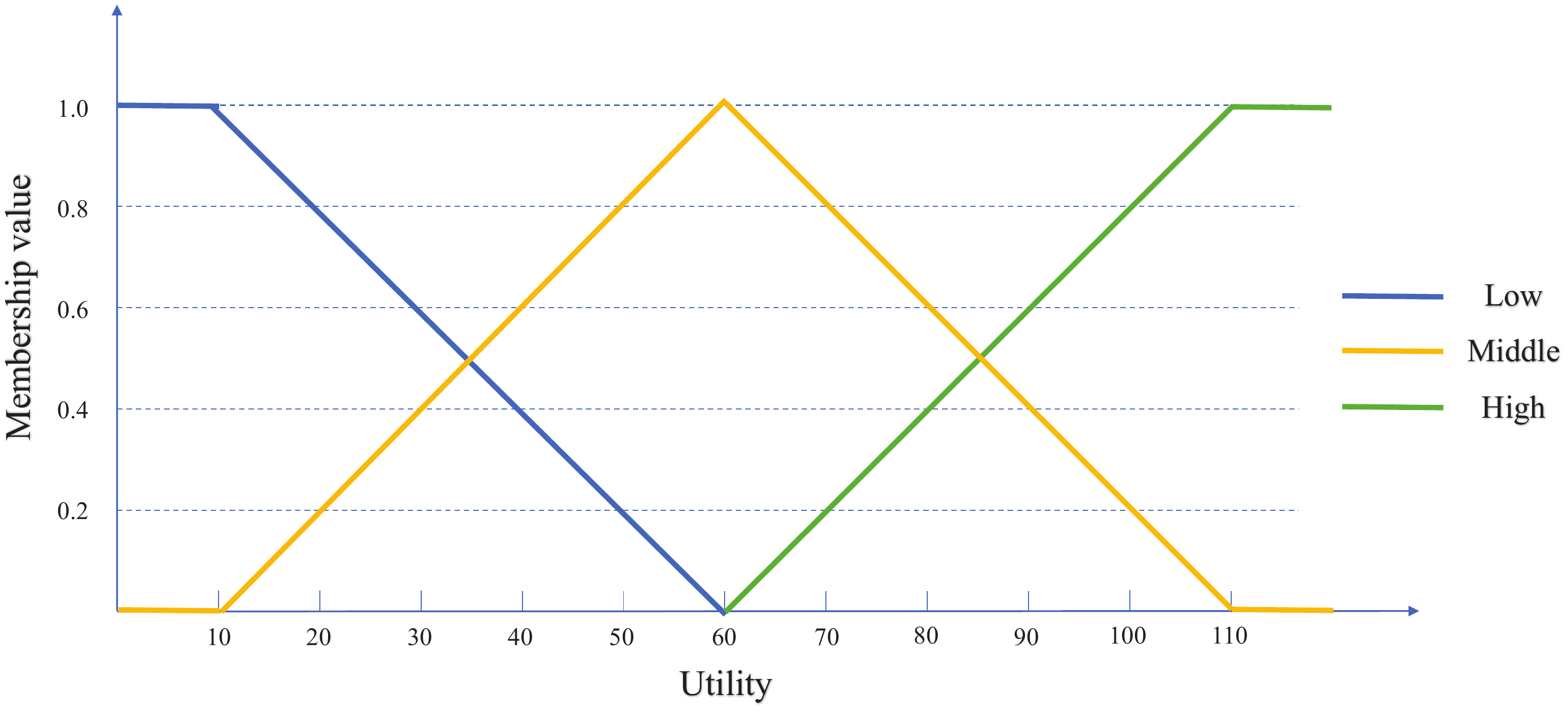}
	\caption{Membership function used in experiments.}
	\label{mf}
\end{figure}

Herein, we present a series of experiments, implemented in Java JDK 1.8, that were conducted to evaluate the performance of the methods under comparison. All programs were executed on a PC with a 3.8 GHz Intel Core i7-10700K CPU and 32 GB RAM running 64-bit Windows 10. For efficiency and feasibility analysis, the state-of-the-art fuzzy PFUS algorithm \cite{lan2013mining} was selected as the baseline. In the experiments, it was assumed that all $q$-items had the same membership function, with the three regions presented in Fig. \ref{mf}, according to a priori knowledge regarding the utility distribution in the datasets.

We verified PFUS and PGFUM on five real datasets and one synthetic dataset. \textit{Yoochoose} \footnote{\url{https://recsys.acm.org/recsys15/challenge/}} and \textit{Kosarak} are two real datasets obtained from a series of click events from an e-commerce site and a Hungarian online news portal, respectively. Some browsing records in \textit{Kosarak} are excessively long, and thus HFUSPs are difficult to mine. Moreover, \textit{Bible}, \textit{Leviathan}, and \textit{Sign} were utilized as linguistic datasets; they were derived from the Bible, the famous novel Leviathan, and sign language utterances of video segments, respectively. In these datasets, each word can be transformed into a digital item. The datasets represent most data types with various features that are typically encountered in real-world scenarios. Except for \textit{Yoochoose}, the datasets are publicly available in the open-source library\footnote{\url{http://www.philippe-fournier-viger.com/spmf/}}. We also generated a more complex synthetic dataset with $q$-sequences. The datasets were widely used in previous studies \cite{gan2020proum,gan2020fast}, and details can be found in Ref. \cite{gan2020proum}.

Details on all databases can be found in Table \ref{features}, where $|D|$ and $|I|$ represent the number of $q$-sequences and different items in the dataset, respectively, $\textit{avg}(S)$/$\textit{max}(S)$ indicate the average/maximum length of the $q$-sequences, \textit{\#Seq} is the average number of $q$-itemsets per $q$-sequence, and \textit{\#Ele} is the average number of $q$-items per $q$-itemset.

\begin{table}[!htbp]
	\caption{Features of datasets}
	\label{features}
	\centering       
	\begin{tabular}{|c|c|c|c|c|c|c|}
		\hline
		\textbf{Dataset} & \textbf{$|\textit{D}|$} & \textbf{$|\textit{I}|$} & \textbf{$\textit{avg}(\textit{S})$} & \textbf{$\textit{max}(\textit{S})$} & \textbf{\textit{\#Seq}} & \textbf{\textit{\#Ele}} \\ \hline 
		\textit{Yoochoose} & 234,300 & 16,004 & 2.25 & 112 & 1.14 & 1.98 \\ \hline     
		\textit{Kosarak} & 10,000 & 10,094 & 8.14 & 608 & 8.14 & 1.00 \\ \hline
		\textit{Bible} & 36,369 & 13,905 & 21.64 & 100 & 21.64 & 1.00 \\ \hline
		\textit{Leviathan} & 5,834 & 9,025 & 33.81 & 100 & 33.81 & 1.00 \\ \hline
		\textit{Sign} & 730 & 267 & 52.00 & 94 & 52.00 & 1.00 \\ \hline
		\textit{Syn40K}	& 40,000 & 7537	& 26.85 & 213 & 6.20 & 4.33 \\ \hline
	\end{tabular}
\end{table}

\subsection{Runtime Analysis}

\begin{figure*}[ht]
	\centering
	\includegraphics[width=1\linewidth]{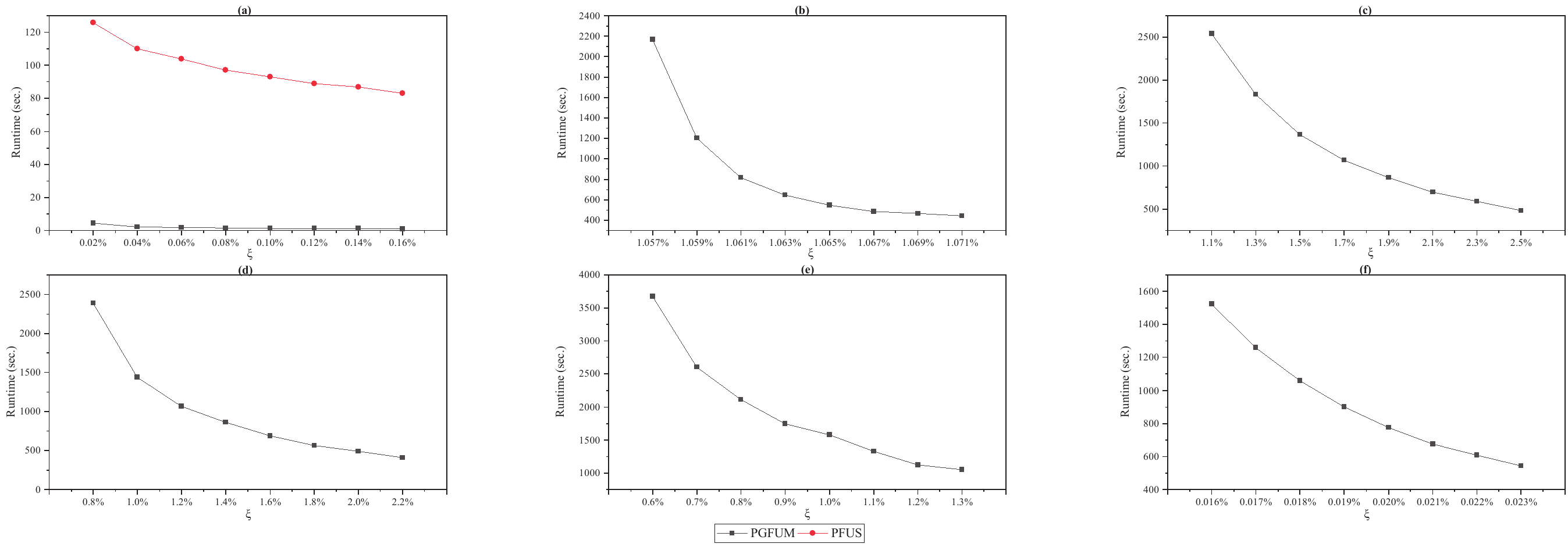}
	\caption{Runtime of compared methods under various minimum fuzzy utility thresholds. (a) Yoochoose. (b) Kosarak. (c) Bible. (d) Leviathan. (e) Sign. (f) Syn40K.}
	\label{runtime}
\end{figure*}

Herein, we are concerned with the runtime of the methods under comparison. Runtime is a crucial performance metric. We conducted extensive experiments using the control variate method under various minimum fuzzy utility thresholds. It is noted that the mining process is forced to terminate once its runtime exceeds 15,000 s, and this is marked using the symbol '/' in the experiments. Details regarding the execution time can be seen in Fig. \ref{runtime}. Evidently, the runtime of the algorithms decreases smoothly as the minimum utility threshold increases. Importantly, the proposed PGFUM significantly outperformed PFUS in all cases. In particular, PFUS could extract the desired HFUSPs only in the \textit{Yoochoose} dataset, where the execution time of PGFUM was better than that of PFUS by several times, on the limited runtime premise. This is because PFUS should search a considerable amount of space and consumed a large amount of time to calculate the utility values of f-sequences, whereas PGFUM effectively pruned the search space by using tight upper bounds. Furthermore, it can be seen that PGFUM required substantially different amounts of time to mine HFUSPs in different datasets owing to their varying features. For example, it required a few seconds in \textit{Yoochoose}, but hundreds of seconds to find the complete set of HFUSPs in other datasets. Thus, it can be observed that there are strong correlations between the average length of the $q$-sequences and the execution time of the algorithms. In conclusion, by adopting the three efficient pruning strategies, PGFUM can effectively discover HFUSPs in a reasonable time.

\subsection{Memory Evaluation}

\begin{figure*}[ht]
	\centering
	\includegraphics[width=1\linewidth]{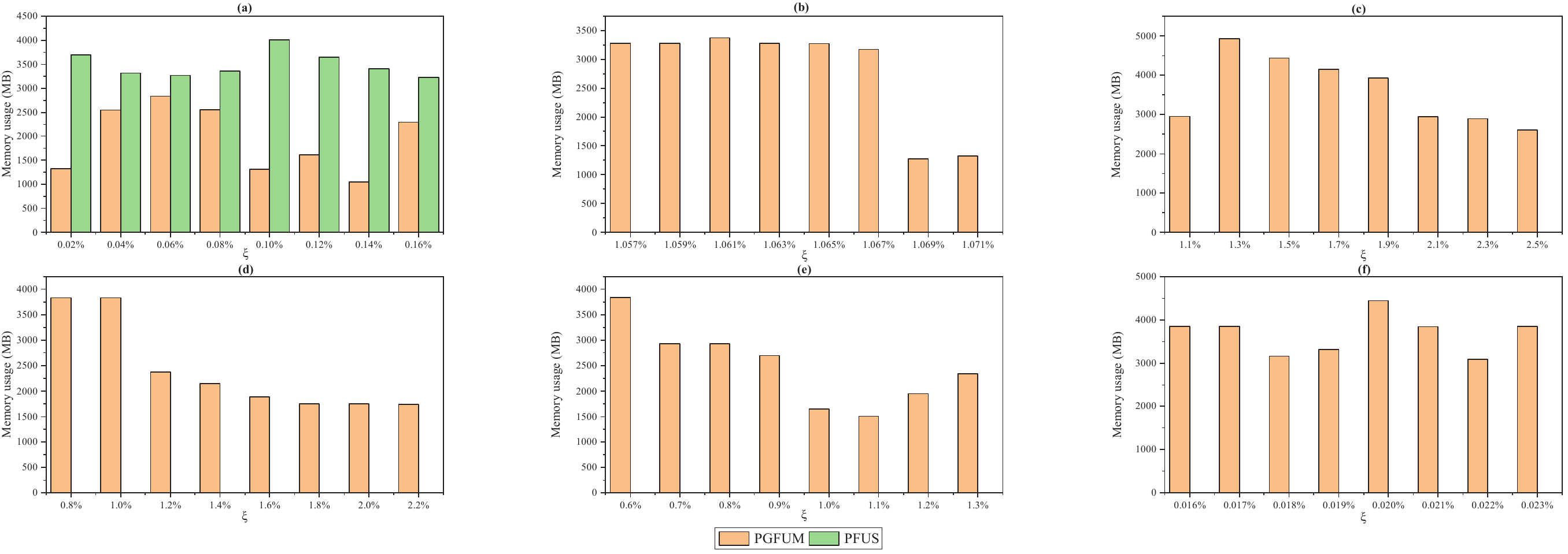}
	\caption{Memory usage of compared approaches by varying different minimum fuzzy utility thresholds. (a) Yoochoose. (b) Kosarak. (c) Bible. (d) Leviathan. (e) Sign. (f) Syn40K.}
	\label{memory}
\end{figure*}

Herein, the memory usage of the two algorithms is compared, as it is a key measure. For the baseline algorithm PFUS, we only present memory consumption on \textit{Yoochosse} because this algorithm cannot be applied to the other datasets, as described in the last subsection. As can be clearly seen in Fig. \ref{memory}, PGFUM significantly outperforms PFUS in terms of memory usage. This is because the efficient pruning strategies used by PGFUM prevent the \textit{FZ-tree} from growing excessively. In particular, the gap of the memory consumption by the two algorithms is the largest when $\xi$ is equal to 0.12\%, whereas the gap closes when $\xi$ is equal to 0.06\% in \textit{Yoochosse}. From a macroscopic point of view, memory usage decreases progressively as $\xi$ increases, whereas runtime increases. However, unlike runtime, memory consumption does not exhibit a stable decline at all times. For instance, it sharply decreases as $\xi$ exceeds 1.067\% and 1.0\% in \textit{Kosarak} and \textit{Leviathan}, respectively. Moreover, PGFUM consumed little memory for storing the projected databases in \textit{Bible} when $\xi$ = 1.1\%. By utilizing the projected database, which is smaller than the original database, PGFUM significantly reduces the scope of scanning. In addition, the fuzzy utility chains store necessary information and ignore information that is irrelevant in the utility calculation; this is beneficial for reducing memory utilization. In conclusion, PGFUM exhibits highly satisfactory performance, as expected.

\begin{figure*}[htbp]
	\centering
	\includegraphics[width=1\linewidth]{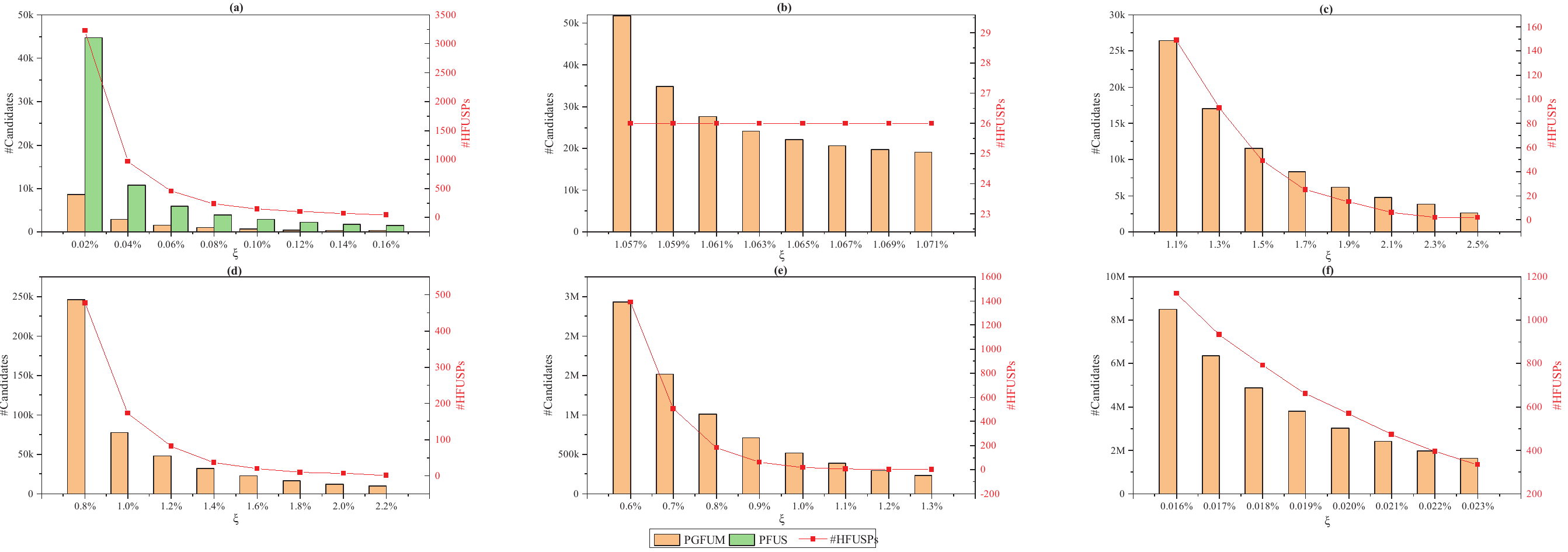}
	\caption{Results of candidates and patterns generated. (a) Yoochoose. (b) Kosarak. (c) Bible. (d) Leviathan. (e) Sign. (f) Syn40K.}
	\label{candidate}
\end{figure*}

\subsection{Candidates and Patterns Analysis}

Herein, we discuss the number of candidates generated by the methods under comparison. This quantity is an important performance measure. In addition, we analyze the number of HFUSPs discovered under various minimum fuzzy utility thresholds. It should be noted that \textit{\#Candidates} and \textit{\#HFUSPs} denote the number of candidates and HFUSPs, respectively. The experimental results are presented in Fig. \ref{candidate}. Even though the \textit{FZ-tree} may theoretically become quite large in complex datasets, in practice, it is relatively small in the proposed PGFUM algorithm. Thus, PGFUM generated fewer candidates than PFUS in the mining process in \textit{Yoochoose}. In the other five datasets, we can safely infer that PFUS should search an excessively large \textit{FZ-tree} and check an excessively large number of candidates, and thus it failed to extract the desired HFUSPs in a reasonable time. It can be concluded that the three proposed pruning strategies efficiently prune the \textit{FZ-tree} structure and reduce the search space as much as possible in all datasets except for \textit{Yoochoose}. More intuitively, an increasing number of candidates are generated as the minimum fuzzy utility threshold increases. This is quite understandable, as the upper bound value of an $f$-sequence is fixed, and thus the pruning strategies may be unable to eliminate some candidates under a larger minimum fuzzy utility threshold. In a dataset, the number of HFUSPs extracted is independent of the methods, but is determined only by the minimum fuzzy utility threshold. In addition, the utility distribution in a dataset has a strong effect on the number of HFUSPs by varying the minimum fuzzy utility threshold. For example, the algorithms discovered 26 HFUSPs in \textit{Kosarak} in all six cases, whereas the number of HFUSPs decreases as the threshold increase. The results shown in Fig. \ref{candidate} indicate the positive effect of pruning strategies on the HFUSP identification performance of PGFUM.

\subsection{Case Studies}

\begin{figure*}[ht]
	\centering
	\includegraphics[width=1\linewidth]{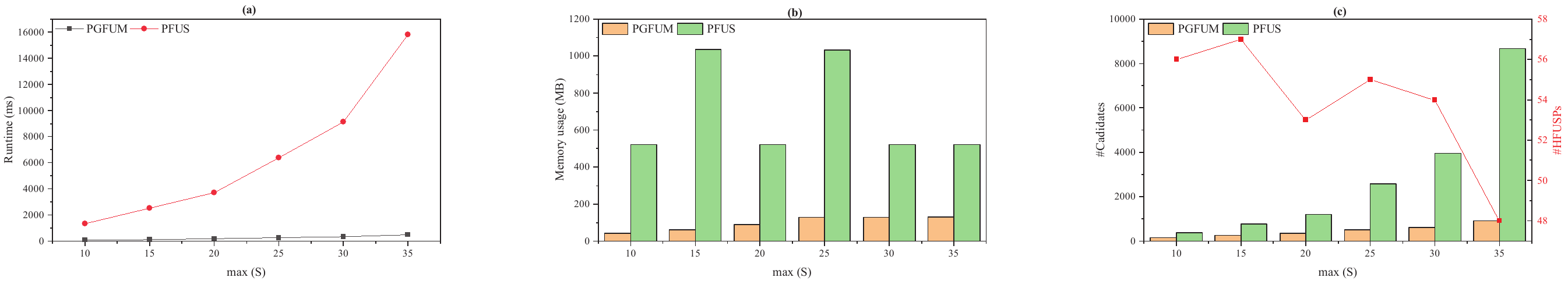}
	\caption{Effectiveness of pruning strategies under various
		\textit{max} $(S)$.}
	\label{case1}
\end{figure*}

\begin{figure*}[ht]
	\centering
	\includegraphics[width=1\linewidth]{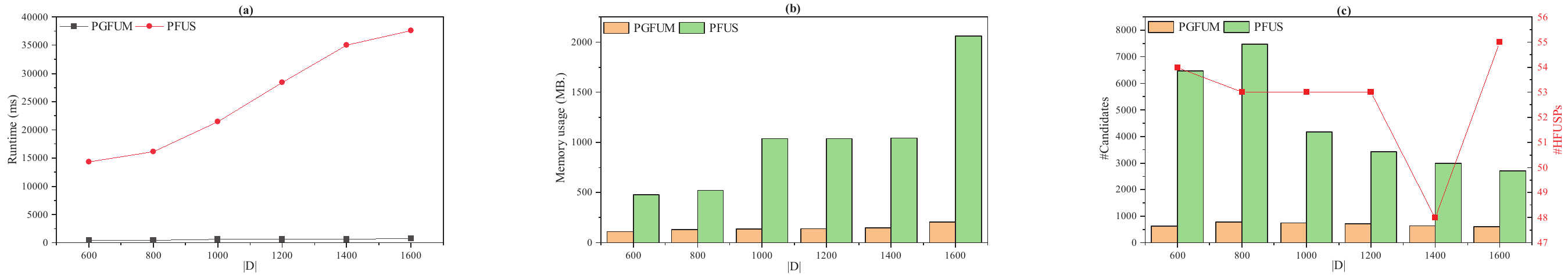}
	\vspace{-0.5cm}
	\caption{Effectiveness of pruning strategies under various $|D|$.}
	\label{case2}
\end{figure*}

As can be clearly observed in Fig. \ref{runtime}, the inefficient algorithm PFUS cannot identify the desired HFUSPs in a reasonable time. We infer that the length and number of $q$-sequences contained in datasets have a significant effect on the performance of mining methods. To evaluate the factors that affect performance, we conducted further mining experiments, where $\xi$ is 1.0\%, on a part of the \textit{Kosarak} dataset. We first generated two sets of datasets derived from \textit{Kosarak}, where the eligible $q$-sequences were extracted, and the others were skipped. For the first set, we generated six datasets by extracting the first 800 $q$-sequences with length not greater than (i.e., \textit{max(S)} equal to) 10, 15, 20, 25, 30, and 35 from \textit{Kosarak}. The experimental results are presented in Fig. \ref{case1} under various values of \textit{max(S)} with $\xi$ = 1.0\%. It is evident that the runtime and the number of generated candidates of the methods under comparison increase as \textit{max(S)} decreases. However, memory usage is not monotonic. For example, the memory consumption of PFUS reaches a peak when \textit{max(S)} is equal to 15 and 25. By contrast, the memory requirement of PGFUM is more stable, and rises slowly as \textit{max(S)} decreases. We note that PGFUM outperforms PFUS in all cases, and the gap between the two algorithms becomes wider when \textit{max(S)} increases. Moreover, the datasets in the second set were generated by extracting from \textit{Kosarak} the first 600, 800, 1000, 1200, 1400, and 1600 $q$-sequences with length not greater than 35. As can be seen in Fig. \ref{case2}, runtime and memory usage increase, as more $q$-sequences should be scanned to calculate utility values. In particular, the memory consumption of PFUS dramatically increases when $|D|$ reaches 1600. Interestingly, the candidates have no connection with $|D|$ because the total utility (i.e., $u(D)$) is larger in datasets containing more $q$-sequences, and $\xi$ is fixed at 1.0\%; thus, $u(D)\times \xi$ becomes larger, and thus more candidates are filtered. In addition, the number of HFUSPs discovered is only influenced by the utility distribution of the dataset. In conclusion, PFUS can only handle these simple datasets, whereas the proposed algorithm PGFUM exhibits excellent performance, particularly when the datasets contain a large number of $q$-sequences, or the $q$-sequences in the dataset are relatively long.

\begin{table}[!ht]
	\scriptsize  
	\caption{Effectiveness of pruning strategies}
	\label{table_effect}
	
	\begin{tabular}{|c|c|c|c|c|p{1cm}<{\centering}|}	
		\hline \textbf{Dataset} & \textbf{Method} & \textbf{PGFUM} & $-\mathbf{_{PPO}}$  & $-\mathbf{_{EUD}}$ & $-\mathbf{_{PES}}$ \\
		\hline
		\multirow{2}{*}{\textit{Yoochoose}} 
		& {Time} & {2.092} & {3.126} & {3.272} & {7.43} \\
		\cline{2-6}
		& {Memory} & {2,542} & {1,275} & {1,040} & {1,283} \\
		\hline
		
		\multirow{2}{*}{\textit{Kosarak}} 
		& {Time} & {645} & {683} & {/} & {706} \\
		\cline{2-6}
		& {Memory} & {3,281} & {2,033} & {/} & {3,499} \\
		\hline
		
		\multirow{2}{*}{\textit{Bible}} 
		& {Time} & {696} & {4,126} & {/} & {794} \\
		\cline{2-6}
		& {Memory} & {2,943} & {3,635} & {/} & {3,256} \\
		\hline
		
		\multirow{2}{*}{\textit{Leviathan}} 
		& {Time} & {689} & {1,281} & {/} & {792} \\
		\cline{2-6}
		& {Memory} & {1,881} & {1,909} & {/} & {4,363} \\
		\hline
		
		\multirow{2}{*}{\textit{Sign}} 
		& {Time} & {1,329} & {1,378} & {/} & {1,536} \\
		\cline{2-6}
		& {Memory} & {1,505} & {4,068} & {/} & {4,063} \\
		\hline
		
		\multirow{2}{*}{\textit{Syn40K}} 
		& {Time} & {777} & {792} & {/} & {1,456} \\
		\cline{2-6}
		& {Memory} & {4,443} & {3,853} & {/} & {3,848} \\			
		\hline
		
	\end{tabular}
\end{table}

\subsection{Effectiveness Analysis}

\begin{figure}[ht]
	\centering
	\includegraphics[trim=0 0 0 0,clip,scale=0.4]{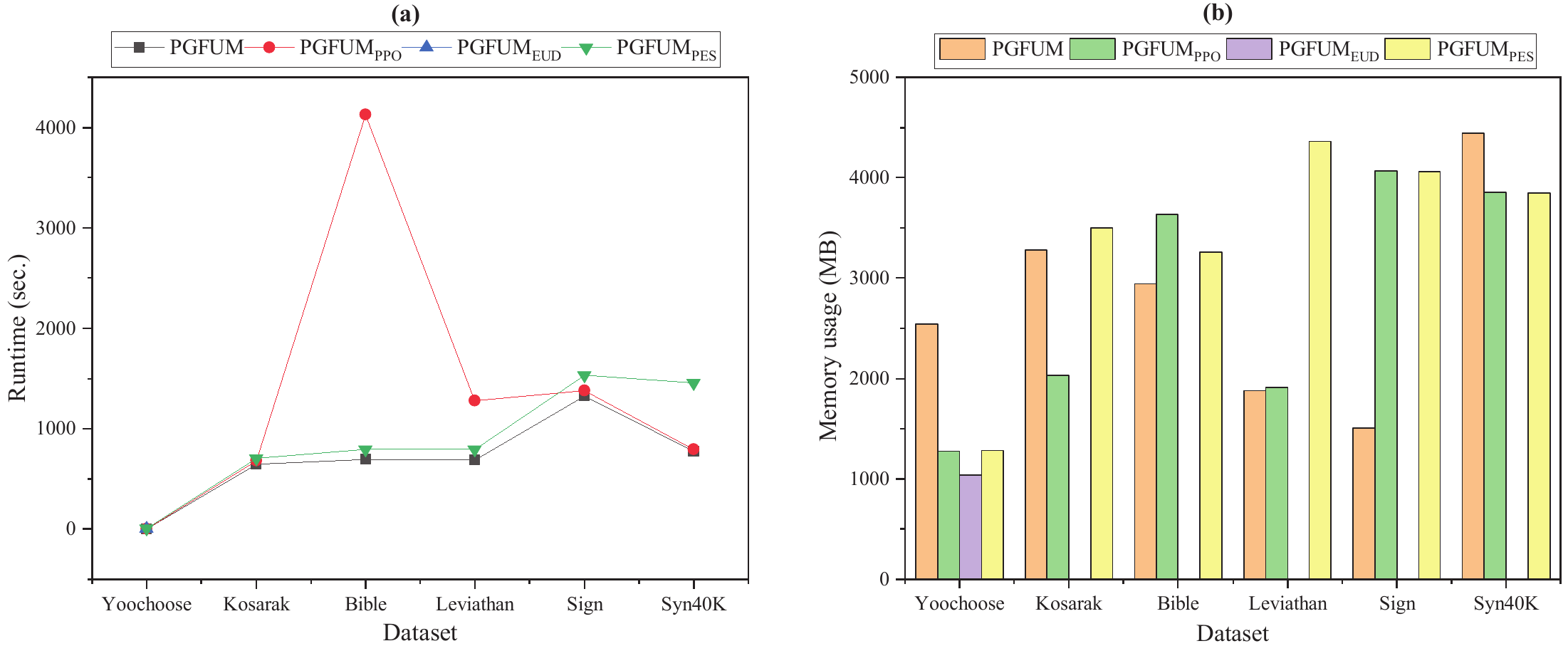}
	\vspace{-0.5cm}
	\caption{Effectiveness of pruning strategies.}
	\label{effectiveness}
\end{figure}

To test the effectiveness of the three proposed pruning strategies, we designed three variants, each without one of the strategies, based on the backbone algorithm PGFUM: $\rm{PGFUM}_{\rm{PPO}}$, $\rm{PGFUM}_{\rm{EUD}}$, and $\rm{PGFUM}_{\rm{PES}}$. We evaluated the runtime and memory usage of PGFUM, and those of the three variant methods on the six datasets with $\xi$ equal to 0.04\%, 1.063\%, 2.1\%, 1.6\%, 1.1\%, and 0.02\%. The performance of these methods is presented in Table \ref{table_effect}, and the results are shown in Fig. \ref{effectiveness}. As can be clearly seen in Table \ref{table_effect}, PGFUM requires the shortest execution time on the six tests, as the efficient data structures and novel strategies greatly to contribute to expediting fuzzy utility calculation and identifying candidates in the database. However, PGFUM did not perform the best in terms of memory usage; it consumed the smallest amount of memory only in \textit{Bible}, \textit{Leviathan}, and \textit{Sign}, that is, the three linguistic datasets. We can conclude that, in terms of memory usage, the proposed PGFUM with the three pruning strategies performs better on datasets in which the average number of $q$items per $q$-itemset is relatively small. Moreover, unlike the others, the variant $\rm{PGFUM}_{\rm{EUD}}$ was unable to discover the desired HFUSPs in a reasonable amount of time in most datasets. This demonstrates that the EUD pruning strategy contributes the most among the three pruning strategies to the efficiency improvement. Moreover, $\rm{PGFUM}_{\rm{PPO}}$ exhibited clearly inferior performance on the three relatively simple linguistic datasets, but its performance was similar to that of PGFUM on other complex datasets. We can also infer that to mine HFUSPs in more complex datasets, the proposed upper bounds \textit{SDFU} and \textit{EIFU} are more suitable than \textit{HFSUUB} for pruning the subtrees of the \textit{FZ-tree}.

\begin{figure*}[!htbp]
	\centering
	\includegraphics[width=1\linewidth]{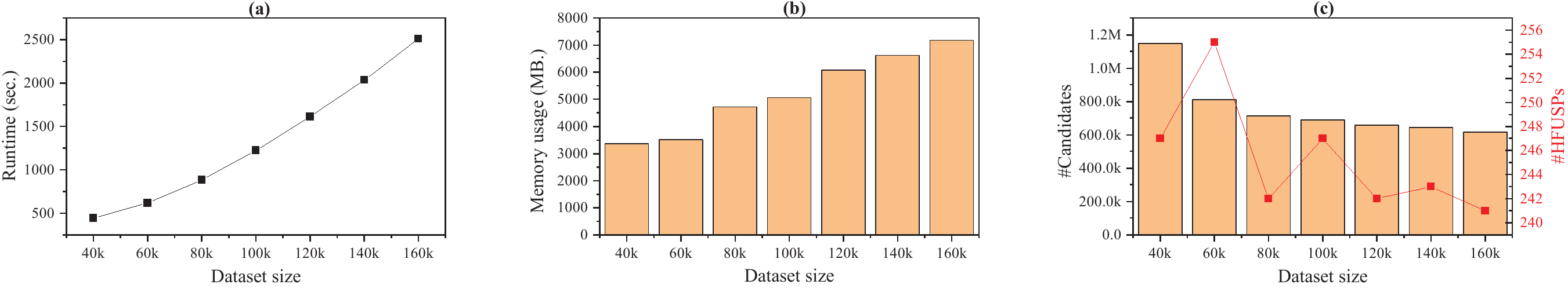}
	\vspace{-0.5cm}
	\caption{Scalability of the compared methods.}
	\label{scalability}
\end{figure*}

\subsection{Scalability}

To evaluate the capability to handle massive data, we conducted a series of experiments on large datasets to test the scalability of PGFUM. The datasets used in the experiments were synthesized with sizes varying from 40K to 160K. We note that we also conducted further experiments using the PFUS method with the same parameter settings on the six datasets. However, PFUS could not discover the desired HFUSPs in any case, as it only adopts one pruning strategy, and this contributes little to the search space reduction. The performance of PGFUM in terms of runtime, memory usage, and number of candidates and patterns is shown in Fig. \ref{scalability}.  It can be seen that runtime and memory usage decrease smoothly as the dataset size increases, whereas the number of generated candidates generated and that of discovered HFUSPs is not proportional to the size of the datasets. For example, PGFUM generated more than one million candidates in the dataset with 40K $q$-sequences, whereas the number of candidates was less than thirty thousand in the dataset with 60K $q$-sequences. In the other datasets, the number of candidates remained stable and was approximately six or seven hundred thousand. PGFUM generated an excessively large number of candidates, but the runtime was small in the smallest dataset, because the projected database is relatively small, and therefore it was not time-consuming to calculate utilities. Evidently, the runtime and memory consumption are almost linearly related to the dataset size, clearly demonstrating that TKUS is well scalable.

\section{Conclusion and Future Studies} 
\label{sec:conclusion}

Fuzzy systems have powerful modeling capabilities of explainability and interpretability. In this study, we investigated explainable fuzzy-theoretic utility mining on multi-sequences and gave a more normative formulation of the problem of FUM on sequences. To handle the issue, we proposed a novel explainable algorithm termed PGFUM, which integrates fuzzy theory and utility mining to achieve human-explainable mining results, to mine HFUSPs on multi-sequences with linguistic meaning for decision making. In particular, two compressed data structures with explainable fuzziness, namely, fuzzy matrix set and fuzzy utility chain, were designed to compress rich information in multi-sequences. Meanwhile, two new upper bounds on the explainable fuzzy utility and three pruning strategies were proposed to substantially reduce the search space and thus expedite the mining process. It was demonstrated that PGFUM not only yields human-explainable mining results that contain the original nature of revealable intelligibility, but also achieves better performance than the state-of-the-art algorithm. 

In future work, attempts will be made to further enhance the execution speed of this approach. For example, we intend to design and implement a distributed version of PGFUM on a cloud computing platform, such as Hadoop and Spark. Furthermore, FUM on sequences does not offer a measure of confidence or probability that an HFUSP may be followed. A better framework that extends the existing method should be developed, guided by discovering high-fuzzy-utility rules. In addition, how to further apply the principle of FUM to other practical applications is also an interesting issue, like the task of mining explainable patterns in uncertain data or stream environments.

\section*{Acknowledgment}

Our gratitude goes to the anonymous reviewers for their careful work and thoughtful suggestions that have helped improve this paper substantially.

\ifCLASSOPTIONcaptionsoff
  \newpage
\fi

\bibliographystyle{IEEEtran}
\bibliography{PGFUM}

\vspace{-1.5cm}
\begin{IEEEbiography}[{\includegraphics[width=1in,height=1.25in,clip,keepaspectratio]{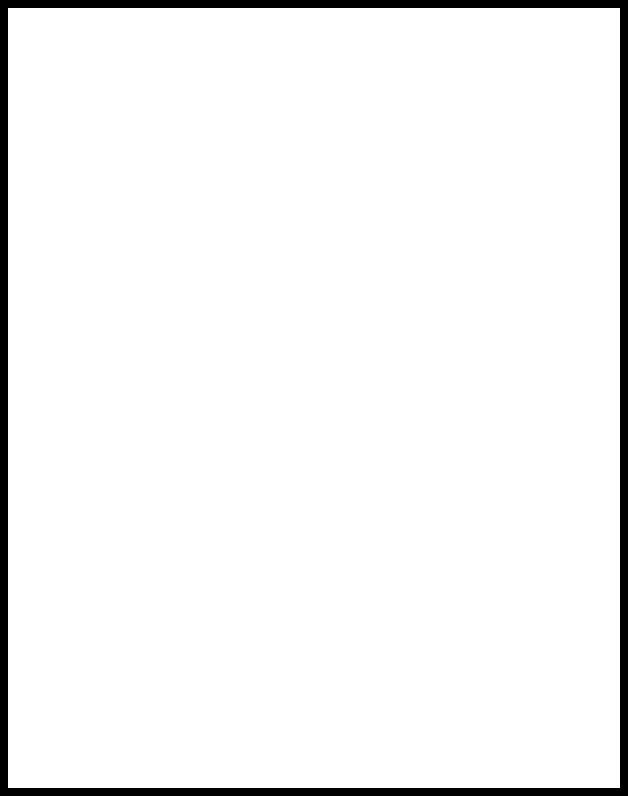}}]{Wensheng Gan  (Member, IEEE)} received the Ph.D. in Computer Science and Technology, Harbin Institute of Technology (Shenzhen), Guangdong, China in 2019. He is currently an Association Professor with the College of Cyber Security, Jinan University, Guangzhou, China.  His research interests include data mining, utility mining, and big data. He has published more than 70 research papers in peer-reviewed journals and international conferences. He is an Associate Editor of \textit{Journal of Internet Technology}. 
\end{IEEEbiography}

\vspace{-1.5cm}
\begin{IEEEbiography}[{\includegraphics[width=1in,height=1.25in,clip,keepaspectratio]{figs/newAuthor.png}}]{Zilin Du (Student Member, IEEE)}
	received the B.S. degree in Computer Science from Huaqiao University, Xiamen, China in 2015. He is a graduate student with the Department of Computer Science and Technology, Harbin Institute of Technology (Shenzhen), Shenzhen, China. His research interests include data mining, machine learning, and artificial intelligence.
\end{IEEEbiography}

\vspace{-1.5cm}
\begin{IEEEbiography}[{\includegraphics[width=1in,height=1.25in,clip,keepaspectratio]{figs/newAuthor.png}}]{Weiping Ding (Senior Member, IEEE)}
	is currently a Professor with the School of Information Science and Technology, Nantong University, Jiangsu, China. He received the Ph.D. degree in computation application from the Nanjing University of Aeronautics and Astronautics (NUAA), Nanjing, China, in 2013. His research interests include fuzzy systems, data mining, and artificial intelligence. He has published more than 80 peer-reviewed journal and conference papers. Dr. Ding serves on the Editorial Advisory Board of \textit{Knowledge-Based Systems} and Editorial Board of \textit{Information Fusion}, \textit{Applied Soft Computing}. 
\end{IEEEbiography}

\vspace{-1.5cm}
\begin{IEEEbiography}[{\includegraphics[width=1in,height=1.25in,clip,keepaspectratio]{figs/newAuthor.png}}]{Chunkai Zhang (Member, IEEE)}
	received the Ph.D. degree from Shanghai Jiaotong University, Shanghai, China, 2001. He is currently an Associate Professor with the Department of Computer Science and Technology, Harbin Institute of Technology (Shenzhen), Shenzhen, China. His current research interests include data mining and cyber security. Dr. Zhang is a Member of IEEE. 
\end{IEEEbiography}

\vspace{-2cm}
\begin{IEEEbiography}[{\includegraphics[width=1in,height=1.25in,clip,keepaspectratio]{figs/newAuthor.png}}]{Han-Chieh Chao (Senior Member, IEEE)}
	has been the president of National Dong Hwa University since February 2016. He received M.S. and Ph.D. degrees in Electrical Engineering from Purdue University in 1989 and 1993, respectively. His research interests include wireless networks, data science, and artificial intelligence. He has published nearly 500 peer-reviewed professional research papers. He is the Editor-in-Chief (EiC) of IET Networks and \textit{Journal of Internet Technology}. Dr. Chao is an IEEE Senior Member and a fellow of IET. 
\end{IEEEbiography}

\end{document}